\documentclass[prd,preprintnumbers,floatfix,aps,notitlepage,nofootinbib,twocolumn,showpacs,amssymb]{revtex4-2}
\usepackage{amsmath,amsfonts,bm}
\usepackage{graphicx}
\usepackage{amssymb}
\usepackage{cancel}
\usepackage{booktabs} 
\usepackage{array}    
\usepackage{amsmath}
\usepackage[colorlinks, linkcolor={red},citecolor={blue}]{hyperref}
\usepackage{xcolor}

\begin{document}
\title{Formation of the Kerr black hole: an exact model}
\author{J Ovalle}
\email[]{jorge.ovalle@physics.slu.cz}
\affiliation{Research Centre for Theoretical Physics and Astrophysics,
	Institute of Physics, Silesian University in Opava, CZ-746 01 Opava,
	Czech Republic}	

\affiliation{Universidad de Tarapac\'a, Avenida Luis Emilio Recabarren 2477, Iquique, Chile
}
\begin{abstract}
		\noindent We present an exact analytical model of axisymmetric gravitational collapse leading to the formation of the Kerr black hole. The model depends only on the total mass ${\cal M}$ and a time dependent rotation parameter $a(v)$, and incorporates both the extremal Kerr black hole and a novel quasi extremal regime without requiring $a\approx{\cal M}$. The evolution develops curvature singularities induced by $a(v)$, which are expected to remain enclosed within a trapped region close to the evolving Kerr radius $r=h(v)$. It also predicts a transient anisotropic exterior curvature decaying as $1/r^2$, which may imprint observable signatures associated with the formation of a rotating black hole. The present construction should therefore be interpreted as an exact analytical description of the final stage of rotational gravitational collapse, immediately preceding the formation of the Kerr black hole.
		\end{abstract} 

\maketitle
%
%
%
\section{Introduction}

\noindent Any realistic stellar system possesses, in addition to its total mass ${\cal M}$ an angular momentum $J$. If such a system undergoes complete gravitational collapse, all the complexities associated with the collapsing matter are expected to be radiated away during a transient stage, ultimately yielding a stationary configuration characterized solely by its mass and angular momentum, in accordance with the uniqueness theorem~\cite{Israel:1967wq,Israel:1967za,Carter:1971zc,Hawking:1971vc,Robinson:1975bv,Heusler:1996jaf,Chrusciel:2012jk}. The fundamental feature of this final state is the inevitable appearance of a spacetime singularity, dressed by an event horizon, in accordance with the Penrose theorem~\cite{Penrose:1964wq,Hawking:1970zqf,Hawking:1973uf} and the weak cosmic censorship conjecture~\cite{Penrose:1969pc}, respectively. The above provides, grosso modo, the classical description of the formation of a rotating black hole (BH), whose spacetime is described by the Kerr solution~\cite{Kerr:1963ud}.

Unlike the spherically symmetric case, for which the Oppenheimer-Snyder (OS) model~\cite{Oppenheimer:1939ue} provides an analytical description of the formation of a Schwarzschild BH~\cite{Christodoulou:1984mz,Joshi:1993zg,Joshi:2001xi,Mena:2004ck,Lasky:2006mg,Joshi:2008zz,Mosani:2020ena,Joshi:2023ugm}, the axisymmetric scenario outlined above lacks an analogous model capable of describing the gravitational collapse and eventual formation of a Kerr BH. Indeed, given its complexity, the prospect of obtaining an exact analytical description appears little more than a chimera. The main obstacle is the absence of an analog of Birkhoff's theorem~\cite{Teukolsky:2014vca}: one cannot simply prescribe a Kerr exterior for the collapsing object, since the evolution of the latter will inevitably modify the former. Consequently, there exists an infinite family of admissible exterior geometries that may describe the exterior of a rotating collapsing configuration. Paradoxically, the greater freedom in choosing the exterior geometry makes the construction of an exact analytical model substantially more difficult. By contrast, in the OS model the exterior spacetime is fixed from the outset, providing a crucial simplification.

In this work, rather than following the OS construction directly, we postulate a single line element describing both the region occupied by the collapsing object and the exterior spacetime. This strategy has proven particularly useful in the spherically symmetric case~\cite{Ovalle:2024wtv,Ovalle:2025pue,Ovalle:2026lxb,Casadio:2026tmd,Lobo:2026dnl,Lobo:2026iuy}, where the formation of the Schwarzschild BH can be described analytically and precisely without resorting to highly idealized matter sources. The same perspective was recently extended to the axisymmetric case in Ref.~\cite{Ovalle:2026ajv}.

The model constructed here is a time-dependent extension of the family of rotating solutions recently reported in Ref.~\cite{Ovalle:2026ajv}. It possesses two defining properties: (i) it depends only on the pair of parameters $\{{\cal M},a\}$ and therefore carries no primary hair; and (ii) it remains as close as possible to the Kerr-Schild class, to which the Kerr solution belongs. As a consequence, we obtain an exact analytical model of axisymmetric gravitational collapse leading to the formation of the Kerr BH, including both the extremal case with $a={\cal M}$ and a novel quasi-extremal regime with $a\not\approx{\cal M}$. Even though the resulting spacetime is intrinsically singular, the singularity is expected to remain enclosed within a trapped region located close to the evolving Kerr radius $r=h(v)$. The construction therefore provides an exact description of the final stage of gravitational collapse, immediately preceding the formation of the Kerr BH, and establishes an analytical framework for investigating its formation in the absence of additional gravitational hairs.

 \section{Revisited Kerr BH}
 \label{sec2}
 \noindent Before constructing a time-dependent extension of the Kerr BH family introduced in Ref.~\cite{Ovalle:2026ajv}, let us briefly review its main properties. Let us start with the Kerr-Schild metric in Boyer-Lindquist coordinates, namely, the Gurses-Gursey metric~\cite{Gurses:1975vu}
\begin{eqnarray}
	\label{metric}
	ds^{2}
	&=&
	-\left[1-\frac{2\,r\,\tilde{m}(r)}{{\rho}^2}\right]
	dt^{2}
	-
	\frac{4\, {a}\, r\,\tilde{m}(r)\, \sin^{2}\theta}{{\rho}^{2}}
	\,dt\,d\phi
	\nonumber
	\\
	&&
	+
	\frac{{\rho}^{2}}{{\Delta}}\,dr^{2}
	+
	{\rho}^{2}\,d\theta^{2}
	+
	\frac{{\Sigma}\, \sin^{2}\theta}{{\rho}^{2}}\,d\phi^{2}
	\ ,
\end{eqnarray}
with
\begin{eqnarray}
	{\rho}^2
	&=&
	r^2+{a}^{2}\cos^{2}\theta\ ,
	\label{f0}
	\\
	{\Delta}
	& = &
	r^2-2\,r\,\tilde{m}(r)
	+{a}^{2}\ ,
	\label{f2}
	\\
	{\Sigma}
	& = &
	\left(r^{2}+{a}^{2}\right)^{2}
	-{\Delta}\, a^2\sin^{2}\theta\ ,
	\label{f3}
\end{eqnarray}
and
\begin{equation}
	\label{spin}
	{a}\,=\,{J}/{\cal M}\ ,
\end{equation}
where ${J}$ is the angular momentum and
\begin{equation}
	\label{cond1}
	{\cal M}\equiv \tilde{m}(h)
\end{equation}
is the Arnowitt-Deser-Misner (ADM) mass, with $r=h$ denoting the event-horizon radius, given by
\begin{equation}
	\label{kerrh}
	h
	=
	{\cal M}+\sqrt{{\cal M}^2-a^2}
\end{equation}
 with
\begin{equation}
	\label{mtransform}
	\tilde{m}(r)=\left\{
	\begin{array}{l}
		m(r)
		\ ,
		\quad
		{\rm for}\
		0< r \leq h
		\\
		\\
	{\cal M}
		\ ,
		\quad
		{\rm for}\
		r>h
		\ .
	\end{array}
	\right.
\end{equation}
Here $m(r)$ denotes the Misner-Sharp mass associated with the corresponding spherically symmetric configuration $a=0$. The classical Kerr solution is recovered from~\eqref{mtransform} by taking $\tilde{m}(r)={\cal M}$ for all $0<r<\infty$.
 
 To ensure smooth continuity of the metric~\eqref{metric} across $r=h$, the mass function must satisfy
 \begin{equation}
 	\label{cond2}
 	m(h)={\cal M}\ ; \qquad m'(h)=0\ ,
 \end{equation}
 where $F(h)\equiv\,F(r)\big\rvert_{r=h}$ for any $F(r)$. Condition~\eqref{cond2} guarantees that the metric is ${\cal C}^{1}$ across the horizon. By imposing higher differentiability across the horizon, one can construct mass functions belonging to the class ${\cal C}^{N}$~\cite{Ovalle:2025pue,Ovalle:2024wtv,Casadio:2024fol,Casadio:2025pun}.  Under these conditions, the interior mass function takes the form
  \begin{eqnarray}
 	\label{minfi}
 	&&	m(r)=\left[\left(\frac{r}{h}\right)^3\,\prod_{i=1}^{N}\frac{n_i+1}{n_i-2}\right.\nonumber\\
 	&&\left.+3(-1)^N\,\sum_{k=1}^{N}\frac{1}{n_k-2}\left(\frac{r}{h}\right)^{n_k+1}\prod_{\substack{i=1\\i\neq k}}^{N}\frac{n_i+1}{n_k-n_i}\right]{\cal M}\ ,
 \end{eqnarray} 
 where $2<n_i\in\mathbb{N}$. For each fixed $N$, the set $n_i=\{n_1, n_2,\ .\ .\ .n_N\}$, labels an infinite family of regular Kerr BHs. Since this configurations depends solely on the parameters $\{{\cal M},\,a\}$, they carry no primary hairs. 
 
The equation $\Delta(r)=0$ in~\eqref{f2} generally admits several roots, whose number is determined by the largest exponent in the set $n_i=\{n_1, n_2,\ .\ .\ .n_N\}$. Nevertheless, only two roots are real within the interval $0<r\leq h$: the event horizon at $r=h$ and an inner (Cauchy) horizon $h_c<h$. The horizon at $r=h$ is therefore generically nondegenerate, becoming degenerate only in the extremal limit $a={\cal M}$, where $h_c=h$. The exceptional case $m(r)={\cal M}(r/h)^3$ instead corresponds to a horizonless mimicker configuration.
 
For illustration, let us consider the simplest cases  $N=1$,\footnote{In this case the term $i\neq\,k$ produces an empty product and therefore evaluates to $1$.} 
and $N=2$, for which the mass function~\eqref{minfi} reduces to 
\begin{equation}
	\label{m1}
	m(r)=\frac{{\cal M}}{(n-2)}\left[\frac{r^3}{h^3}\left(n+1\right)-3\left(\frac{r}{h}\right)^{n+1}\right]\ ;\,\,\,n>2\ ,
\end{equation}
\begin{eqnarray}
	\label{m2}
	m(r)=&&\frac{r}{h}\left[\frac{(n+1)(l+1)}{(n-2)(l-2)}\left(\frac{r}{h}\right)^2+\frac{3\,(l+1)}{(n-2)(n-l)}\left(\frac{r}{h}\right)^n\right.
	\nonumber\\
	&&\left.+\frac{3\,(n+1)}{(l-2)(l-n)}\left(\frac{r}{h}\right)^l \right]{\cal M}\ ;\,\,\,l>n>2\,\in\mathbb{N}\ ,
\end{eqnarray}
respectively. In the limit $a=0$, both solutions reduce to the regular spherically symmetric configurations originally obtained in Ref.~\cite{Ovalle:2024wtv} and later analyzed in Refs.~\cite{Casadio:2025pun,Ovalle:2026lxb,Casadio:2026tmd,Lobo:2026dnl}.

Regarding the scalar curvature, it takes the form
\begin{eqnarray}
	\label{Rinfi}
	\hspace*{-6mm}	&&	R(r,\theta)=\frac{6\cal M}{\rho^2\,h}\left[4\,\prod_{i=1}^{N}\frac{n_i+1}{n_i-2}\left(\frac{r}{h}\right)^2\right.\nonumber\\
	\hspace*{-6mm}	&&\left.+(-1)^N\,\sum_{k=1}^{N}\frac{(n_k+1)(n_k+2)}{(n_k-2)}\left(\frac{r}{h}\right)^{n_k}\prod_{\substack{i=1\\i\neq k}}^{N}\frac{n_i+1}{n_k-n_i}\right]
\end{eqnarray} 
with the property that $R(h,\theta)=0$ for $N>1$. The remaining curvature invariants, such as the Kretschmann scalar, have more involved expressions. Nevertheless, they show that the ring singularity $\rho=0$ never develops provided $n_i>2$ for all $i$. Relaxing the condition $n_i>2$ to $n_i \in [-2,2]$, however, yields BHs with integrable singularities~\cite{Lukash:2013ts,Ovalle:2023vvu,Arrechea:2025fkk}. In particular, whenever $n_i=-1$ for some $i$, the Kerr geometry is recovered, thereby providing a unified framework for investigating the emergence of the Kerr BH during gravitational collapse

We conclude by highlighting three important features of the metric~\eqref{metric} with mass function~\eqref{minfi}: (i) it depends only on the charges $\{{\cal M},a\}$ and therefore carries no primary hair; (ii) it possesses a single inner (Cauchy) horizon $h_c$; and (iii) it admits a quasi-extremal regime in which $h_c\rightarrow{h}$, even when $a$ is far from the extremal Kerr value ${\cal M}$.

\section{Evolution}
\label{sec3}
\noindent To analyze the evolution of the geometry described by Eq.~\eqref{metric}, with mass profile $\tilde{m}$ defined by Eq.~\eqref{mtransform} and interior mass function $m(r)$ given by~\eqref{minfi}, we first extend the configuration to a time dependent setting. As a preliminary step, the stationary metric~\eqref{metric} is rewritten in Eddington-Finkelstein form by introducing the ingoing null coordinate $v$ together with the shifted azimuthal coordinate $\bar{\phi}$ through~\cite{Chandrasekhar:1983}
\begin{equation}
	\label{newcoor}
		v=t+\int\frac{r^2+a^2}{\Delta}dr
		\ ;\,\,\,\,\bar{\phi}=\phi+\int\frac{a}{\Delta}dr\ ,
\end{equation}
We then replace the static mass function by its time-dependent counterpart,
\begin{equation}
	\label{promotting}
	\tilde{m}(r)\rightarrow\,\tilde{m}(v,r)\ ,
\end{equation}
and similarly promote the spin parameter in Eq.~\eqref{spin} to
\begin{equation}
	\label{a(v)}
a\rightarrow\,{a(v)}\,=\,{J(v)}/{\cal M}\ .
\end{equation}
These steps lead to the following Eddington-Finkelstein-like metric (dropping bars in $\bar{\phi}$)
\begin{eqnarray}
	\label{EF-metric}
	&&ds^{2}
	=
	-\left[1-\frac{2\,r\,\tilde{m}}{{\rho}^2}\right]
	dv^{2}+2\,dv\,dr	+
	{\rho}^{2}\,d\theta^{2}
	\nonumber
	\\
	&&
	-2a \sin^{2}\theta\,dr\,d\phi-
	\frac{4\, {a}\, r\,\tilde{m}\, \sin^{2}\theta}{{\rho}^{2}}
	\,dv\,d\phi
	+
	\frac{{\Sigma}\, \sin^{2}\theta}{{\rho}^{2}}\,d\phi^{2} ,\nonumber\\
\end{eqnarray}
where $(v,r,\theta,\phi)\in\mathbb{R}\times(0,\infty)\times{S}^2$. The metric functions in~\eqref{f0}-\eqref{f3} are now generalized according to
\begin{eqnarray}
	&&{\rho(v,r,\theta)}^2
	=
	r^2+{a(v)}^{2}\cos^{2}\theta\ ,
	\label{f0v}
	\\
	&&{\Delta(v,r)}
	= 
	r^2-2\,r\,\tilde{m}(v,r)
	+{a(v)}^{2}\ ,
	\label{f2v}
	\\
	&&{\Sigma(v,r,\theta)}
	= 
	\left[r^{2}+{a(v)}^{2}\right]^{2}
	-{\Delta(v,r)}\, a(v)^2\sin^{2}\theta\ ,
	\label{f3v}
\end{eqnarray}
subject to the condition
\begin{equation}
	\label{cond12}
	{\cal M}\equiv \tilde{m}(v,h(v))\neq\,	{\cal M}(v)\ ,
\end{equation}
where $h(v)$ denotes the evolving Kerr-like radius
\begin{equation}
	\label{kerrh2}
	h(v)
	=
	{\cal M}+\sqrt{{\cal M}^2-a(v)^2}\ ,
\end{equation}
which, by construction, satisfies $\Delta(v,h(v))=0$. 

Due to the time dependence of the rotation parameter~\eqref{a(v)}, the metric~\eqref{EF-metric} no longer belongs to the Kerr-Schild subclass,
\begin{equation}
	\label{nonKS}
	g_{\mu\nu}\neq\eta_{\mu\nu}+\frac{2r\tilde{m}}{\rho^2}l_\mu l_\nu\ ,
\end{equation}
with $l_\mu=\delta_\mu^{\,v}-a\sin^2\theta\delta_\mu^{\,\phi}$. A direct analysis shows that the spacetime is, in general, of Petrov type I~\cite{Stephani:2003tm}.

A profile for $\tilde{m}(v,r)$ satisfying the condition~\eqref{cond12} is given by
\begin{equation}
	\label{mtransform2}
	\tilde{m}(v,r)=\left\{
	\begin{array}{l}
		m(v,r)
		\ ,
		\quad
		{\rm for}\
		0< r \leq h(v)
		\\
		\\
		{\cal M}
		\ ,
		\quad
		{\rm for}\
		r\geq h(v)
		\ ,
	\end{array}
	\right.
\end{equation} 
showing that the total mass ${\cal M}$ of the configuration is always confined within the region bounded by the hypersurface $r=h(v)$. It is important to stress that the exterior region $r>h(v)$ is not stationary, since the line element~\eqref{EF-metric} depends explicitly on $a=a(v)$. Consequently, the exterior geometry is not Kerr, consistently with the absence of a rotational counterpart of Birkhoff's theorem in the axisymmetric case. The Kerr solution is recovered from Eq.~\eqref{EF-metric} in the limit
 \begin{equation}
 	\label{KerrLimit}
 	\begin{array}{l}
 		\tilde{m}(v,r)\rightarrow{\cal M}\ ,\\[2mm]
 		a(v)\rightarrow a\ ,
 	\end{array}
 \end{equation}
 namely, when the interior mass function $m(v,r)$ in Eq.~\eqref{mtransform2} approaches ${\cal M}$ throughout the domain $0<r<h(v)$ and the rotation parameter becomes time independent, a situation expected to occur in the asymptotic limit $v\to\infty$. This shows that, during the evolution, the purely material contributions $\sim{\tilde{m}}$ and the purely kinetic contributions $\sim{a(v)}$, which arise independently in the non Kerr-Schild geometry~\eqref{nonKS} [see also Eqs.~\eqref{R-EF} and~\eqref{R-EF2}], must be coupled in order to recover the Kerr limit~\eqref{KerrLimit}, as implemented in Eq.~\eqref{n3(t)}.

Regarding the mass function~\eqref{mtransform2} for the region $r\leq h(v)$, it is given by
\begin{eqnarray}
	\label{minfi-v}
	&&	m(v,r)=\left[\left(\frac{r}{h(v)}\right)^3\,\prod_{i=1}^{N}\frac{n_i+1}{n_i-2}+3(-1)^N\times\right.\nonumber\\
	&&\left.\sum_{k=1}^{N}\frac{1}{n_k-2}\left(\frac{r}{h(v)}\right)^{n_k+1}\prod_{\substack{i=1\\i\neq k}}^{N}\frac{n_i+1}{n_k-n_i}\right]{\cal M}\ ,
\end{eqnarray}
where $2<n_i\in\mathbb{N}$. It belongs to the class ${\cal C}^{N}$~\cite{Ovalle:2025pue,Ovalle:2024wtv,Casadio:2024fol,Casadio:2025pun}, since it satisfies 
~\eqref{cond12} together with
\begin{equation}
	\label{cond-n}
	\frac{d^{n} m(v,r)}{dr^{n}}\bigg|_{r=h}=0\ ,
\end{equation}
for every $1\le n\le N$. As in the stationary case~\eqref{minfi}, each fixed value of $N$ defines a family specified by the set $n_i=\{n_1, n_2,\ .\ .\ .n_N\}$, yielding an infinite collection of time dependent rotating configurations entirely determined by the pair $\{{\cal M},\,a\}$, and therefore introducing no additional parameters.

Now, by differentiating the evolving Kerr radius $h(v)$ in~\eqref{kerrh2}, we obtain
\begin{equation}
	\label{kerrh3}
	\dot{h}=-\frac{a\dot{a}}{\sqrt{{\cal M}^2-a^2}}=-\frac{a\dot{a}}{h-{\cal M}}\ ,
\end{equation}
which explicitly reveals two possible regimes: collapse ($\dot{h}<0$) with increasing spin parameter ($\dot{a}>0$), and expansion ($\dot{h}>0$) with decreasing spin parameter ($\dot{a}<0$). Moreover, since
\begin{equation}
	\label{Delta-v}
\Delta(v,h(v))=0\ ,
\end{equation}
we can carry out an analysis fully analogous to that of the Kerr BH, leading to
\begin{equation}
	\delta{\cal M}=\frac{\kappa}{8\pi}\delta{A}+\Omega\delta{J}\ ,
\end{equation}
a relation formally identical to the first law of BH mechanics~\cite{Bardeen:1973gs}. For $\delta{\cal M}=0$, it reduces to
\begin{equation}
	\label{area}
	\delta{A(v)}=-\frac{8\pi\Omega(v)}{\kappa(v)}\delta{J(v)}\ ,
\end{equation}
where $\{A,\Omega,\kappa\}$ denote the area, angular velocity, and surface gravity associated with $h(v)$, respectively.

\subsection{Singularities}

\noindent A point of central importance regarding the development of singularities and the eventual formation of the Kerr BH concerns the scalar curvature $R$ for the metricc~\eqref{EF-metric},
\begin{eqnarray}
	\label{R-EF}
	&&R=\frac{2}{\rho^2}(2\tilde{m}'+r\tilde{m}'')+\frac{2r}{\rho^6}\left[2\rho^2+\left(\rho^2-4r^2\right)\sin^{2}\theta\right]a\dot{a}\nonumber\\
	&&-\frac{2}{\rho^4}\left(2\rho^2-r^2\right)\sin^{2}\theta{a}\ddot{a}-\frac{\sin^{2}\theta}{2\rho^6}\left[3\rho^4+8r^2\left(\rho^2-r^2\right)\right]\dot{a}^2\nonumber\\
\end{eqnarray}
which contains a contribution from the mass function $\tilde{m}(v,r)$ and a purely kinetic contribution generated by ${a(v)}$. We note that $R\neq{0}$ even when $\tilde{m}=0$, in agreement with Eq.~\eqref{nonKS}. At $\theta=\pi/2$, Eq.~\eqref{R-EF} reduces to
\begin{eqnarray}
	\label{R-EF2}
	&&R=\frac{2}{r^2}(2m'+rm'')-\left(2 a\ddot{a}+\frac{3}{2}\dot{a}^2\right)\frac{1}{r^2}-\frac{2a\dot{a}}{r^3}
\end{eqnarray}
revealing that the ring singularity persists throughout the evolution whenever $a=a(v)$. In particular, the presence of the non-integrable term $\sim{r^{-3}}$ is especially problematic. To illustrate this, notice that one might be tempted to seek a transformation that removes the singular contributions arising from the kinetic sector in Eq.~\eqref{R-EF2}. For instance, one could require
\begin{equation}
	\label{Rsafe}
	\tilde{m}\rightarrow\bar{m}\quad\Rightarrow\quad{R}\rightarrow\bar{R}=4\Lambda\ ,
\end{equation}
but this cannot be achieved without introducing a mass function containing a logarithmic singularity, as a direct consequence of the term $\sim{r^{-3}}$ in Eq.~\eqref{R-EF2}. Such a singularity inevitably reappears in the Kretschmann scalar ${\cal K}$
\begin{equation}
	\label{msingular}
	\bar{m}\sim{a}\dot{a}\log(r/h)\quad\Rightarrow\quad{\cal K}\sim\left[a\dot{a}\log(r/h)\right]^2/r^6\ .
\end{equation}
Although such transformations may be useful in the context of integrable singularities, in the present work we do not consider regularizations of the scalar curvature $R$ and instead focus exclusively on the formation of the Kerr geometry.

From the above analysis, we conclude that the metric~\eqref{EF-metric} is intrinsically singular whenever the rotation parameter becomes time dependent, $a=a(v)$. However, thanks to the mass profile in Eq.~\eqref{mtransform2}, whose interior mass function $m(v,r)$ is given by~\eqref{minfi-v}, the equation $\Delta(v,r)=0$ admits two roots for any value of $N$, satisfying $0<h_c(v)<h(v)$, just as in the stationary case~\eqref{f2}. This strongly suggests that the singularity associated with Eq.~\eqref{R-EF2} remains enclosed within a trapped region throughout the evolution and is therefore not expected to become naked. A detailed determination of the corresponding trapped surface~\cite{Hayward:1993wb,Ashtekar:2003hk}, expected to lie close to $h(v)$ but not to be spherically symmetric~\cite{Senovilla:2014ika,BenAchour:2025vur}, constitutes a highly nontrivial problem that lies beyond the scope of the present work.

We conclude by noting that, for $r\gg{a}$, namely, far from the collapsing system, the curvature~\eqref{R-EF} behaves as
\begin{equation}
	\label{Rfar}
	R\sim-\left(2 a\ddot{a}+\frac{3}{2}\dot{a}^2\right)\frac{\sin^2\theta}{r^2}	\quad
	{\rm for}\
	r\gg a\ ,
\end{equation}
indicating that during the collapse the exterior geometry develops a transient anisotropic curvature extending far from the collapsing object, which may imprint observable signatures associated with the formation of a rotating BH.

Since our main interest lies in the formation of the Kerr BH, we focus on the evolution toward total collapse, which is analyzed through the following two scenarios.
%
%
%
\begin{table*}
	\caption{Metric function $\Delta(v,r)$ for the case $N=2$ in Eq.~\eqref{minfi-v} [see Eq.~\eqref{m2}] for different choices of $\{n,l\}$. The collapse begins from a regular Schwarzschild (RS) configuration, proceeds through a transient singular Kerr phase, and settles into extremal (E), quasi-extremal (QE) or mimicker (Mi) Kerr stationary configurations.
		\label{tab}}
	\begin{ruledtabular}
		\begin{tabular}{ c c c c c }
			$\{n,\,l\}\ ;\{a_0,\,a_f\}$ & Initial ($a_{0}=0$)&Transient singular [$a(v)$] &Final & $\dot{h}_c$
			\\
			\hline\hline
			$\{2<n<l\}\ ;\{0,\,{\cal M}\}$
			&
			RS, $h=2{\cal M}$& RK, $h(v)
			=
			{\cal M}+\sqrt{{\cal M}^2-a(v)^2}$ &EK: $h_c=h={\cal M}$ & Bounce 
			\\
			\hline
			$\{2<n<<l\}\ ;\{0,\,{\cal M}\}$
			&
			RS, $h=2{\cal M}$& RK, $	h(v)
			=
			{\cal M}+\sqrt{{\cal M}^2-a(v)^2}$  &EK: $h_c=h={\cal M}$  &$\dot{h}_c<0$ 
			\\
			\hline
			$\{2<<n<<l\}\ ;\{0,\,{\cal M}\}$
			&
			QES, $h_c\sim\,h=2{\cal M}$& QEK,	$h_c(v)\sim{h(v)}
			=
			{\cal M}+\sqrt{{\cal M}^2-a(v)^2}$&EK: $h_c=h={\cal M}$   &$\dot{h}_c<0$ 
			\\ 
			\hline
			$\{2<<n<<l\}\ ;\{0,\,\not\approx{\cal M}\}$ & 
			QES, $h_c\sim\,h=2{\cal M}$&QEK, $h_c(v)\sim{h(v)}
			=
			{\cal M}+\sqrt{{\cal M}^2-a(v)^2}$ & QEK: $h_c\sim\,h\not\approx{\cal M}$  & $\dot{h}_c<0$ 
			\\  
			\hline
			$\{n\to\infty,l\to\infty\}\ ;\{0,\,{\cal M}\}$ & 
			S-Mi, $h=2{\cal M}$
			& K-Mi, $h(v)
			=
			{\cal M}+\sqrt{{\cal M}^2-a(v)^2}$&EK-Mi: $h={\cal M}$   & $\nexists$  
			\\ 
		\end{tabular}
	\end{ruledtabular}
\end{table*}
\begin{figure*}
	\centering
	\includegraphics[width=0.325\textwidth]{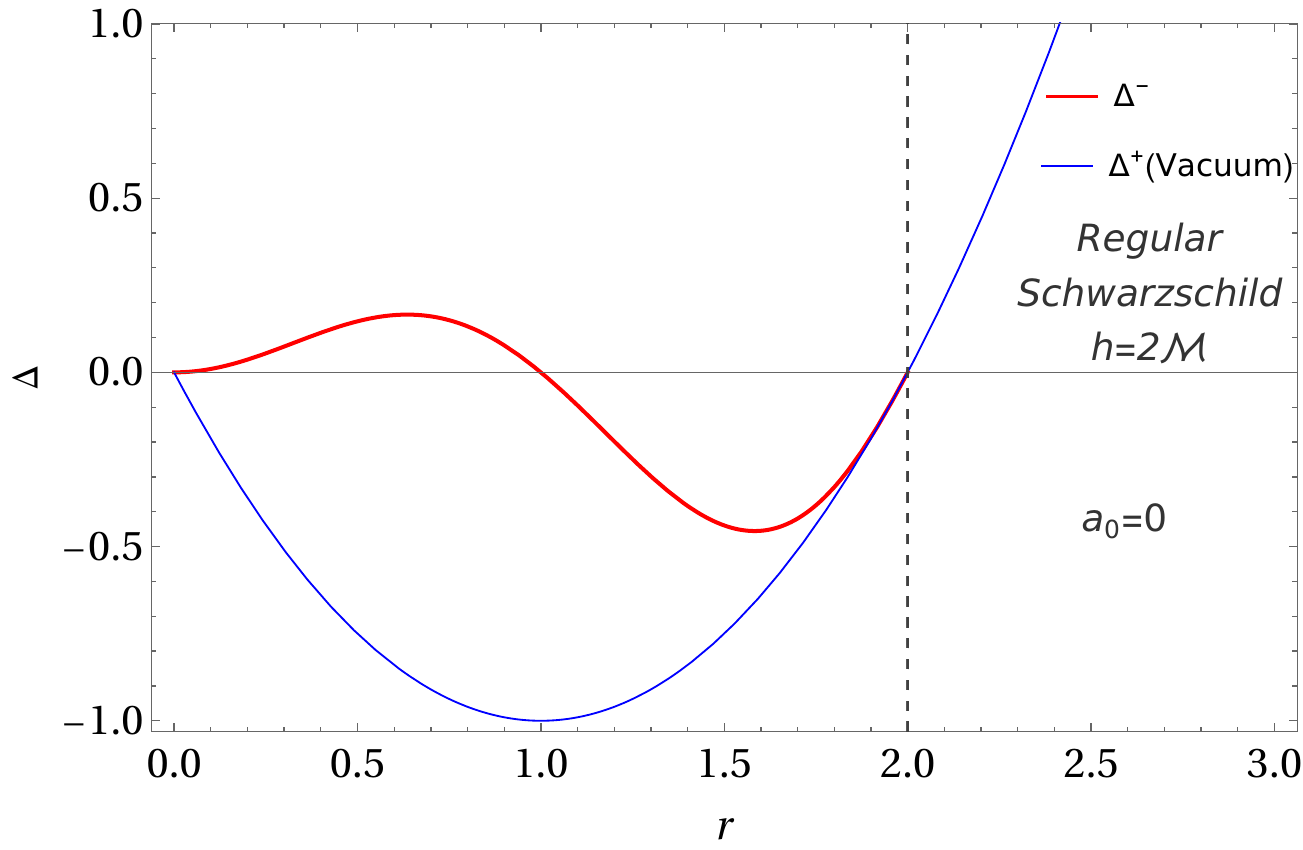} \
	\includegraphics[width=0.325\textwidth]{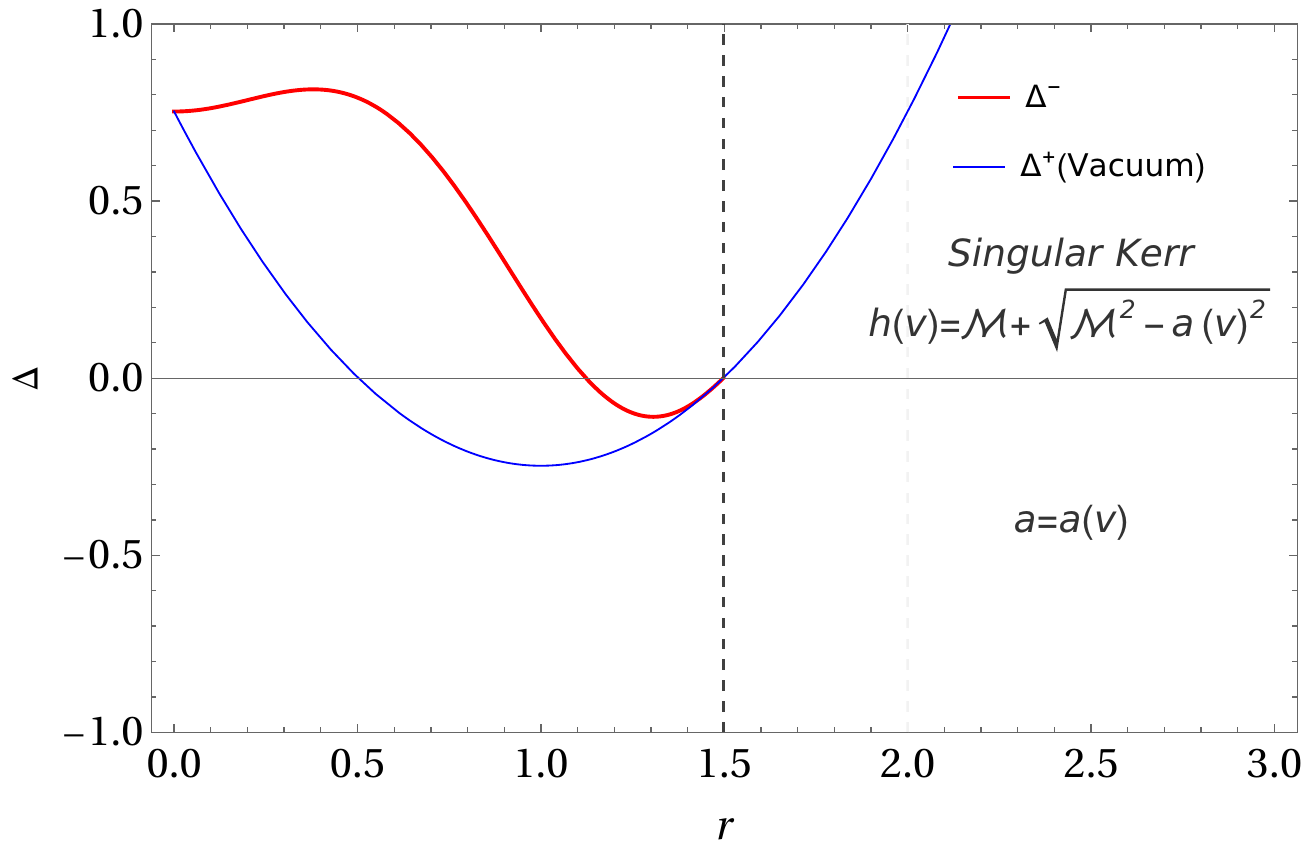} \
	\includegraphics[width=0.325\textwidth]{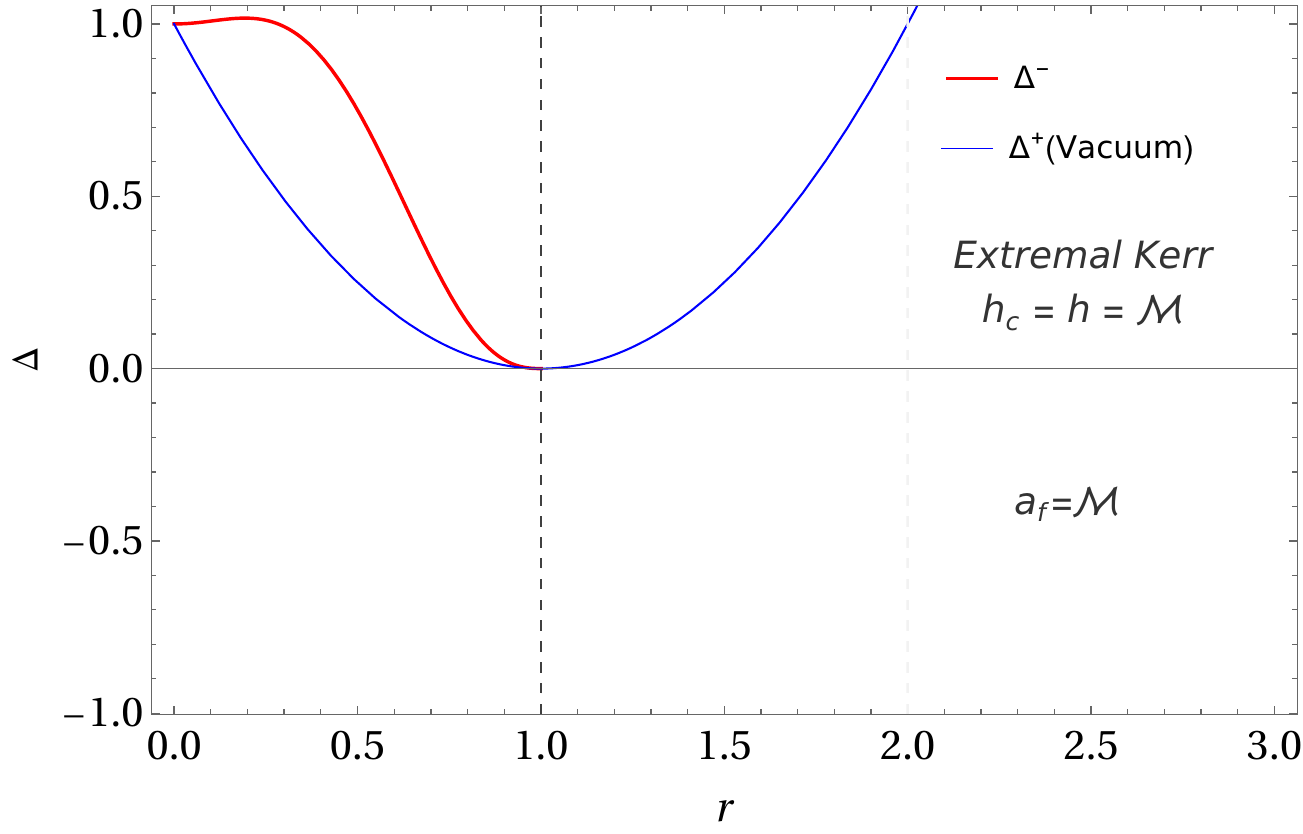} \
	\includegraphics[width=0.325\textwidth]{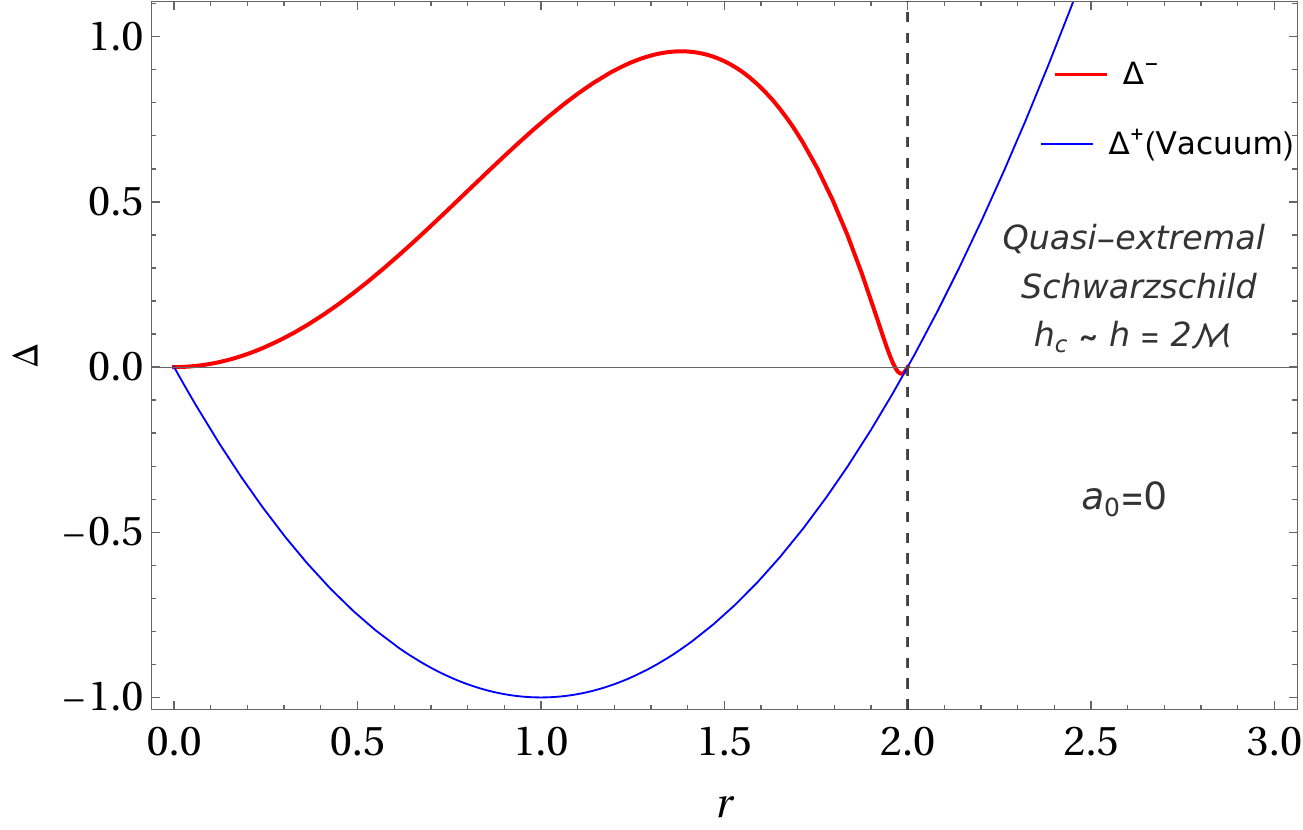} \
	\includegraphics[width=0.325\textwidth]{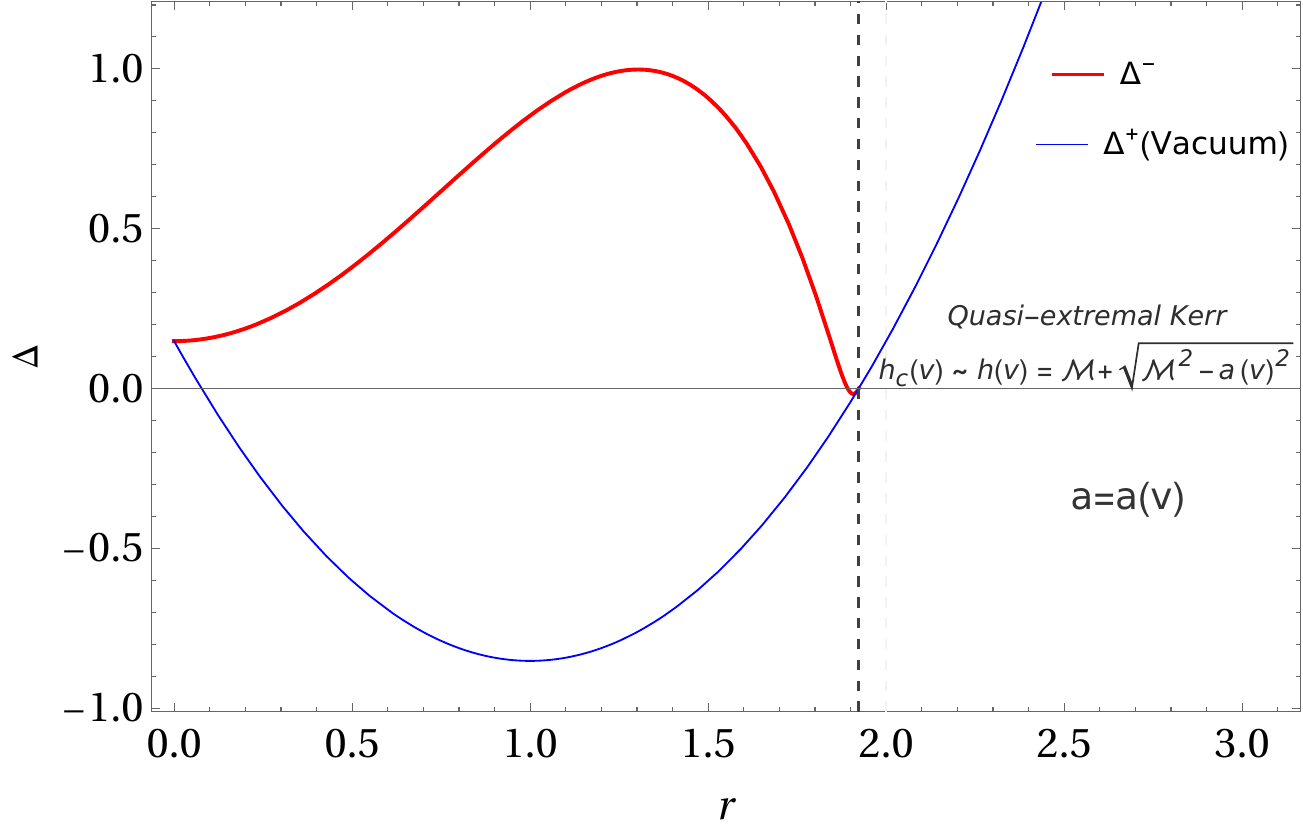} \
	\includegraphics[width=0.325\textwidth]{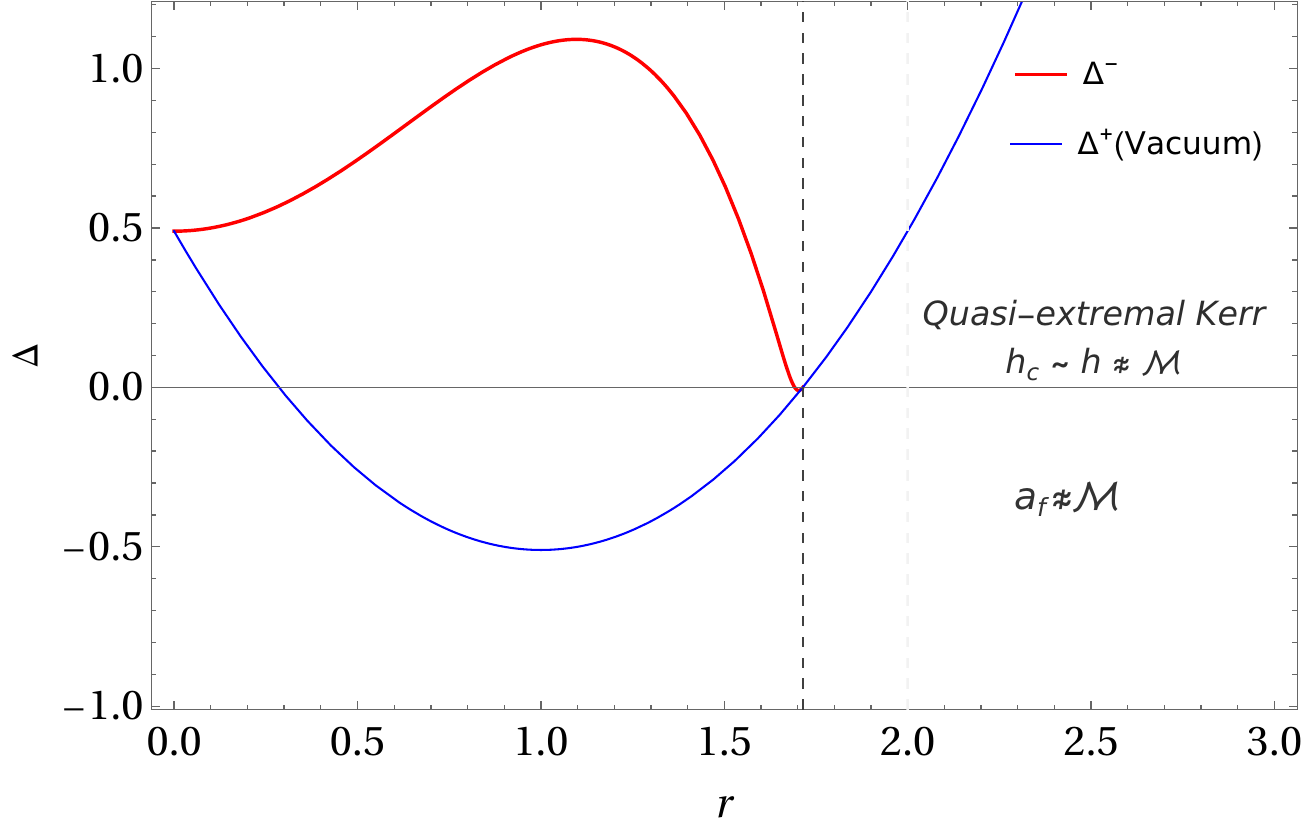} \
	\caption{\footnotesize Metric function $\Delta(v,r)$ for $N=2$ in Eq.~\eqref{minfi-v} [see Eqs.~\eqref{m2}] with $n=3$, $l=4$ (top panels) and $n=100$, $l=200$, $a_f=0.7{\cal M}$ (bottom panels). The spacetime evolves from a regular spherically symmetric configuration (left panels), passes through a transient regime (central panels), and approaches the extremal and quasi-extremal configurations (right panels), respectively. The evolving Kerr radius is located at $h(v)= {\cal M}+\sqrt{{\cal M}^2-a(v)^2}$. The radial coordinate $r$ is measured in units of ${\cal M}$. }
	\label{fig1}
\end{figure*}
\subsection{Case $a=a(v)$ only}
\label{3-A}
\noindent In this case, the evolution is driven exclusively by the unavoidable variation of the angular momentum $a=a(v)$. Let us first analyze the evolution of the mass function~\eqref{mtransform2}, 
\begin{equation}
	\dot{\tilde{m}}=\dot{m}=\frac{\partial{m}}{\partial{h}}\dot{h}+\frac{\partial{m}}{\partial{r}}\frac{dr}{dv}\ ,
\end{equation}
which can be expressed as
\begin{eqnarray}
	\label{mdot1}
	&&	\dot{m}(v,r)=-3{\cal M}\left[\left(\frac{r}{h(v)}\right)^3\,\prod_{i=1}^{N}\frac{n_i+1}{n_i-2}+(-1)^N\times\right.\nonumber\\
	&&\left.\,\sum_{k=1}^{N}\frac{n_k+1}{n_k-2}\left(\frac{r}{h(v)}\right)^{n_k+1}\prod_{\substack{i=1\\i\neq k}}^{N}\frac{n_i+1}{n_k-n_i}\right]\frac{\dot{h}(v)}{h(v)}+m'\dot{r}.\,\,\,\,\,
\end{eqnarray} 
where $\dot{F}(v,r)\equiv{dF}/dv$ and ${F'}(v,r)\equiv{\partial F}/{\partial r}$ for any $F(v,r)$. The expression inside brackets is strictly positive throughout the region $0<r<h(v)$ for any admissible set $\{n_i\}$, with $m'>0$ throughout the same region. Therefore, $\dot{m}(v,r)>0$ whenever $\dot{h}<0$ and $\dot{r}<0$, corresponding to collapse.

Our analysis shows that the configurations described above undergo a progressive contraction accompanied by an increase in angular velocity during collapse. Since this process persists as long as collapsing matter remains present, a natural question concerns the stage at which the evolution may terminate, possibly leading to a stationary regular configuration, assuming such a final state exists. Clarifying this issue requires determining the admissible range of $a(v)$, extending from $a=0$, corresponding to the regular Schwarzschild BH, up to the maximal value $a={\cal M}$ associated with the extremal Kerr BH, which is expected to represent the final stationary state, or a configuration arbitrarily close to it.

We now have all the necessary ingredients to explore the line-element~\eqref{EF-metric} and its evolution toward extremal configurations within the framework of the mass function~\eqref{mtransform2}. To carry out this analysis, we introduce the following smooth interpolating function
\begin{equation}
	\label{a(t)}
	a(v)=\frac{a_0+a_f}{2}+\frac{a_f-a_0}{2}\tanh(\omega{v})\ .
\end{equation}
where $a_0\equiv\,a(-\infty) < a_f\equiv\,a(+\infty)$ characterize the initial and final states, respectively, while $\tau\equiv\omega^{-1}$ sets the characteristic transition timescale. In particular, the interpolation between a regular Schwarzschild BH ($a_0=0$) and an extremal Kerr BH ($a_f={\cal M}$) is described by
\begin{equation}
	\label{a(t)2}
	a(v)=\frac{\cal M}{2}\left[1+\tanh(\omega v)\right]\ ,
\end{equation}
while the interpolation between a regular Schwarzschild BH ($a_0=0$) and an quasi-extremal Kerr BH with $a_f\not\approx{\cal M}$, characterized by $h_c\sim\,h\not\approx{\cal M}$, is described by
\begin{equation}
	\label{a(t)3}
	a(v)=\frac{a_f}{2}\left[1+\tanh(\omega v)\right]\quad \text{for} \quad n_i \to \infty\ \forall\,i\ .
\end{equation}
The corresponding results are summarized in Table~\ref{tab} and displayed in Fig.~\ref{fig1}.

We conclude by emphasizing that, within this framework, the transient singularities disappear once the system reaches the stationary configuration, where the regularity condition $n_i>2$ is satisfied for all $i$. Whether all such regular end states remain stable under generic perturbations~\cite{Penrose:1968,Poisson:1989zz,Ori:1991zz} lies beyond the scope of the present work.

%
%
%
\begin{figure*}
	\centering
	\includegraphics[width=0.325\textwidth]{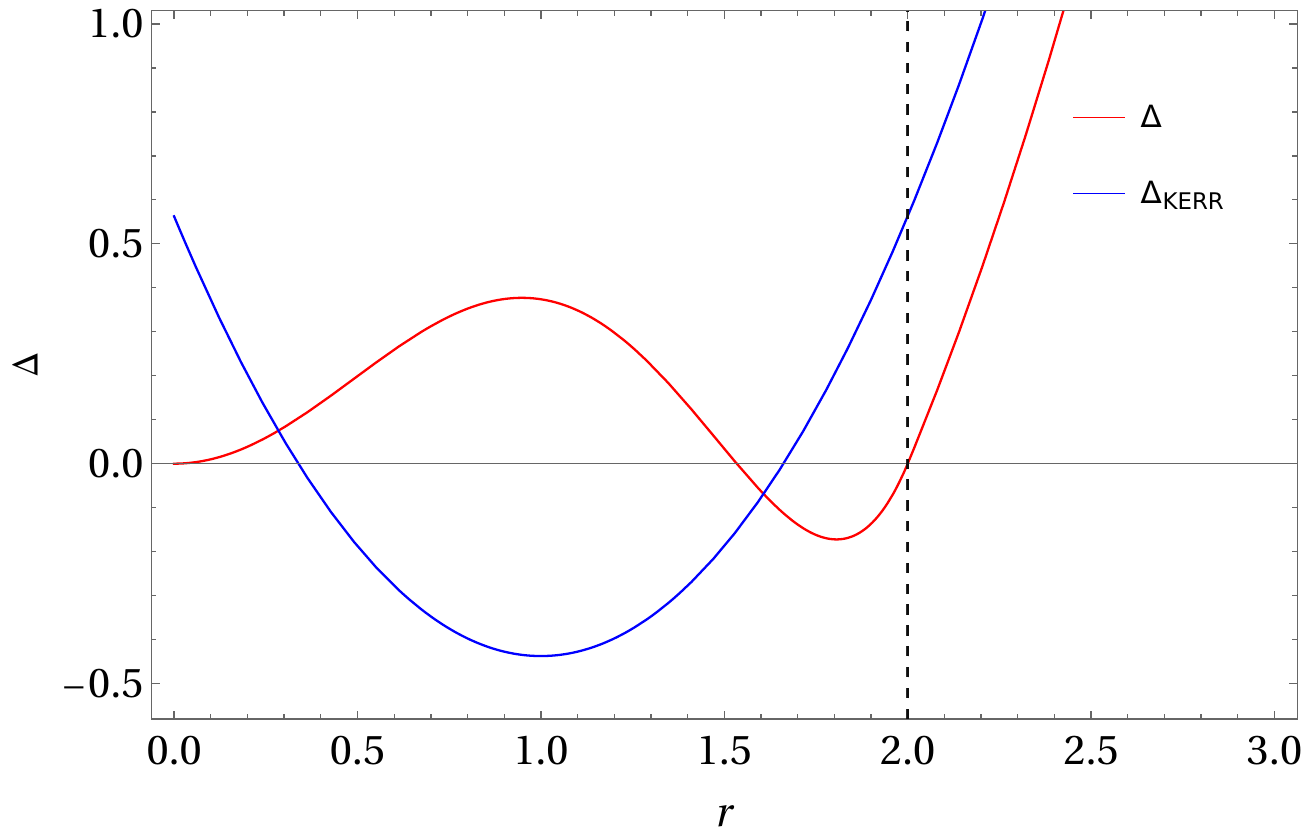} \
	\includegraphics[width=0.325\textwidth]{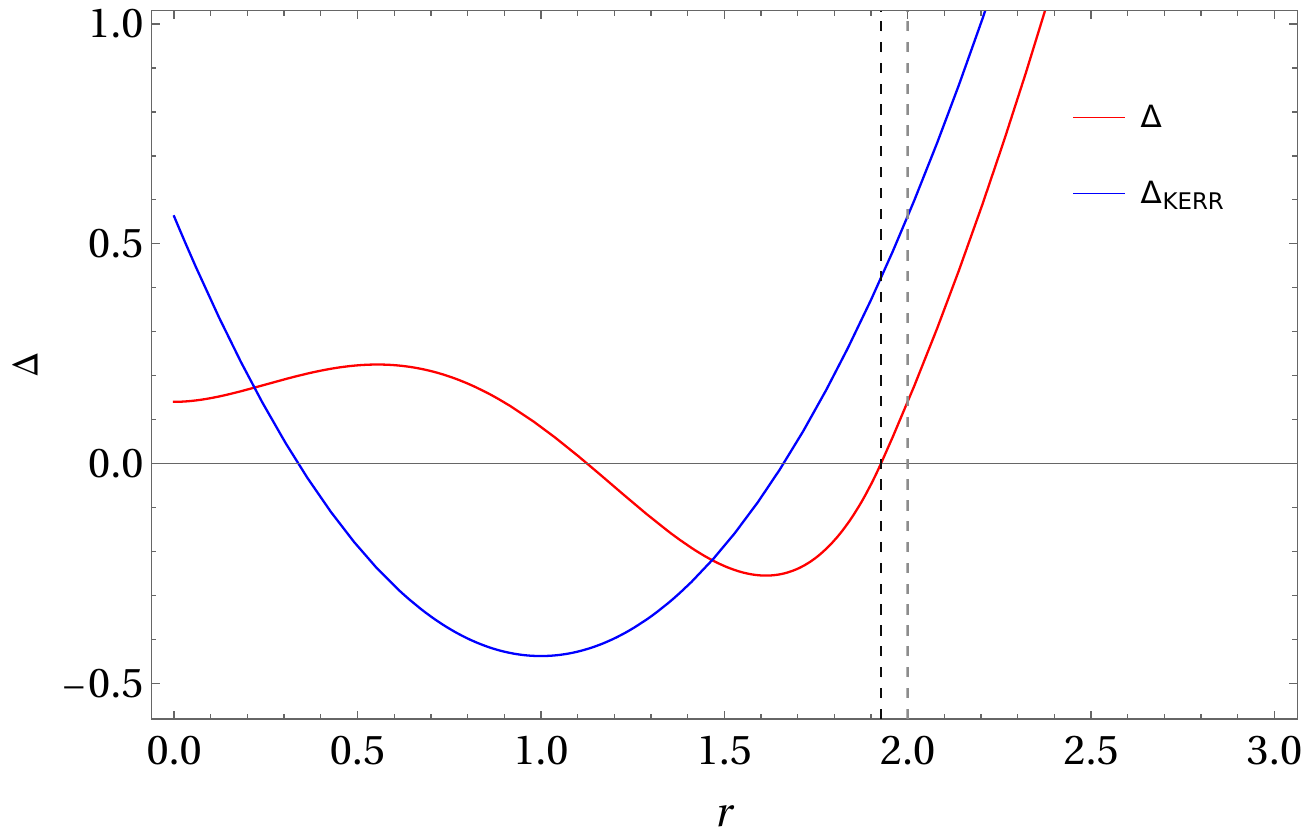} \
	\includegraphics[width=0.325\textwidth]{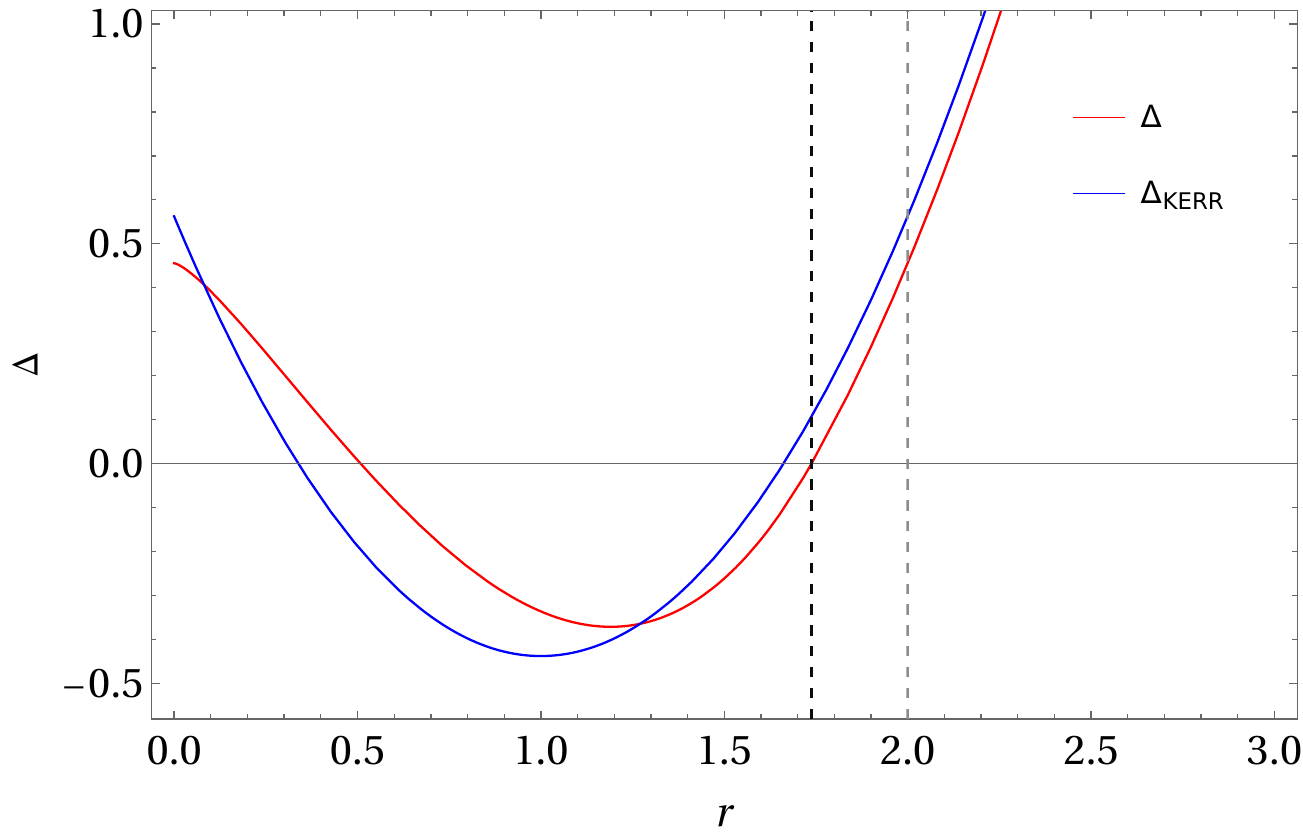} \
	\caption{\footnotesize Metric function $\Delta(v,r)$ in Eq.~\eqref{Delta1} [see also Eqs.~\eqref{a-Schw} and~\eqref{n-Kerr}] at three stages of the evolution prior to the Kerr configuration, shown in blue. The spacetime evolves from an initial regular, spherically symmetric state (left panels) with $\alpha=3$, passes through a transient regime (central and right panels), and ultimately reaches the Kerr configuration with $a_f=0.75{\cal M}$. During the evolution, the Kerr radius $h(v)= {\cal M}+\sqrt{{\cal M}^2-a(v)^2}$ and the inner horizon $h_c(v)$ move toward the center. The frequency is $\omega=0.1$, and the radial coordinate $r$ is measured in units of ${\cal M}$. }
	\label{fig3}
\end{figure*}
%
%
\subsection{Case $a=a(v)$ and $n=n(v)$}
\label{3-B}
\noindent While the previous case is useful for exploring the formation of extremal and quasi-extremal configurations within a simple and fully analytical axially symmetric setting, it lacks a crucial feature: it does not allow us to investigate the formation of the Kerr BH, which represents the primary goal of our analysis. Let us recall that the singular Kerr solution is recovered in the limit $n_i=-1$ for any $i$ in the expression~\eqref{minfi-v}. Reaching this limit requires, besides the variation of $a$, a simultaneous evolution of $n_i$. This leads to the following form of the mass function
\begin{eqnarray}
	\label{minfi-v2}
	&&	m(v,r)=\left[\left(\frac{r}{h(v)}\right)^3\,\prod_{i=1}^{N}\frac{n_i(v)+1}{n_i(v)-2}+3(-1)^N\times\right.\nonumber\\
	&&\left.\sum_{k=1}^{N}\frac{1}{n_k(v)-2}\left(\frac{r}{h(v)}\right)^{n_k(v)+1}\prod_{\substack{i=1\\i\neq k}}^{N}\frac{n_i(v)+1}{n_k(v)-n_i(v)}\right]{\cal M}.\nonumber\\ 
\end{eqnarray}
An inspection of~\eqref{minfi-v2} shows that, in this second scenario, in addition to the singularities associated with the kinetic contribution in the curvature~\eqref{R-EF2}, new singularities appear whenever $n_k(v)=n_i(v)$ for any pair of indices. To avoid this behavior, the functions $n_i(v)$ must evolve without intersections. This type of singularity was first identified in Ref.~\cite{Ovalle:2025pue} for the spherically symmetric case, in connection with the internal geometry of the Schwarzschild BH. To exclude these singularities, we impose an initial ordering at $v=v_0$, namely,
\begin{equation}
	\label{order}
	n_1(v_0)<n_2(v_0)<. . . <n_N(v_0)\ ,
\end{equation}
which must be preserved for all $v$ through the constraint
\begin{equation}
	\label{order2}
	\dot{n}_i(v)\leq\dot{n}_j(v)\ ,\quad\forall\,\, i<j\ .
\end{equation}
[The reversed ordering similarly requires $\dot{n}_i(v)\geq\dot{n}_j(v)$.] The conditions~\eqref{order} and~\eqref{order2} ensure that, throughout the evolution, the singularity remains confined to $r=0$ and enclosed by the roots $\{h_c(v),h(v)\}$, which satisfy $0<h_c(v)<h(v)$.

On the other hand, the resulting expression for $\dot{m}(v,r)$ 
\begin{equation}
	\dot{m}=\left(\frac{\partial\,m}{\partial\,h}\right)\dot{h}+\sum_{i=1}^{N}\left(\frac{\partial\,m}{\partial\,n_i}\right)\dot{n}_i+m'\dot{r}
\end{equation}
is more involved than the one in~\eqref{mdot1}. Nevertheless, a similar analysis shows that, under the conditions~\eqref{order} and~\eqref{order2}, $\dot{m}(v,r)>0$ whenever $\dot{h}<0$ and $\dot{n}_i<0$ for all $i$. Consequently, as in the previous case, the collapsing matter leads to an increase in the angular momentum, and the relation~\eqref{area} remains valid in this more general scenario.

Unlike the previous case, where the final configuration admits the limiting value $a=\mathcal{M}$, there is no lower bound on the functions $n_i(v)$. As a result, the evolution inevitably drives at least one of them to the critical value $n_i(v)=2$, where the contribution of the mass function $m(v,r)$ to the curvature~\eqref{R-EF2} becomes singular.

%
%
%
\begin{figure*}
	\centering
	\includegraphics[width=0.325\textwidth]{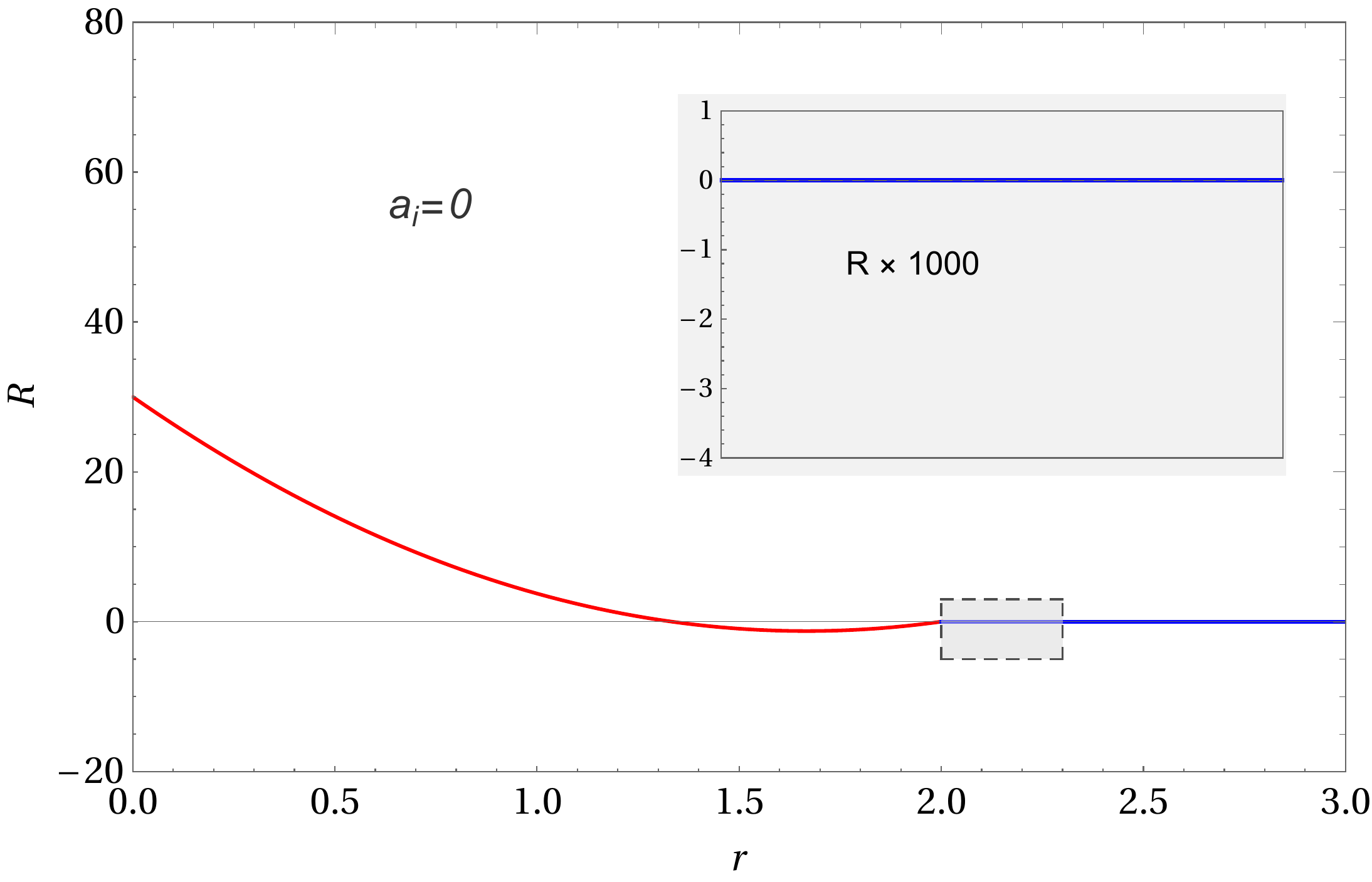} \
	\includegraphics[width=0.325\textwidth]{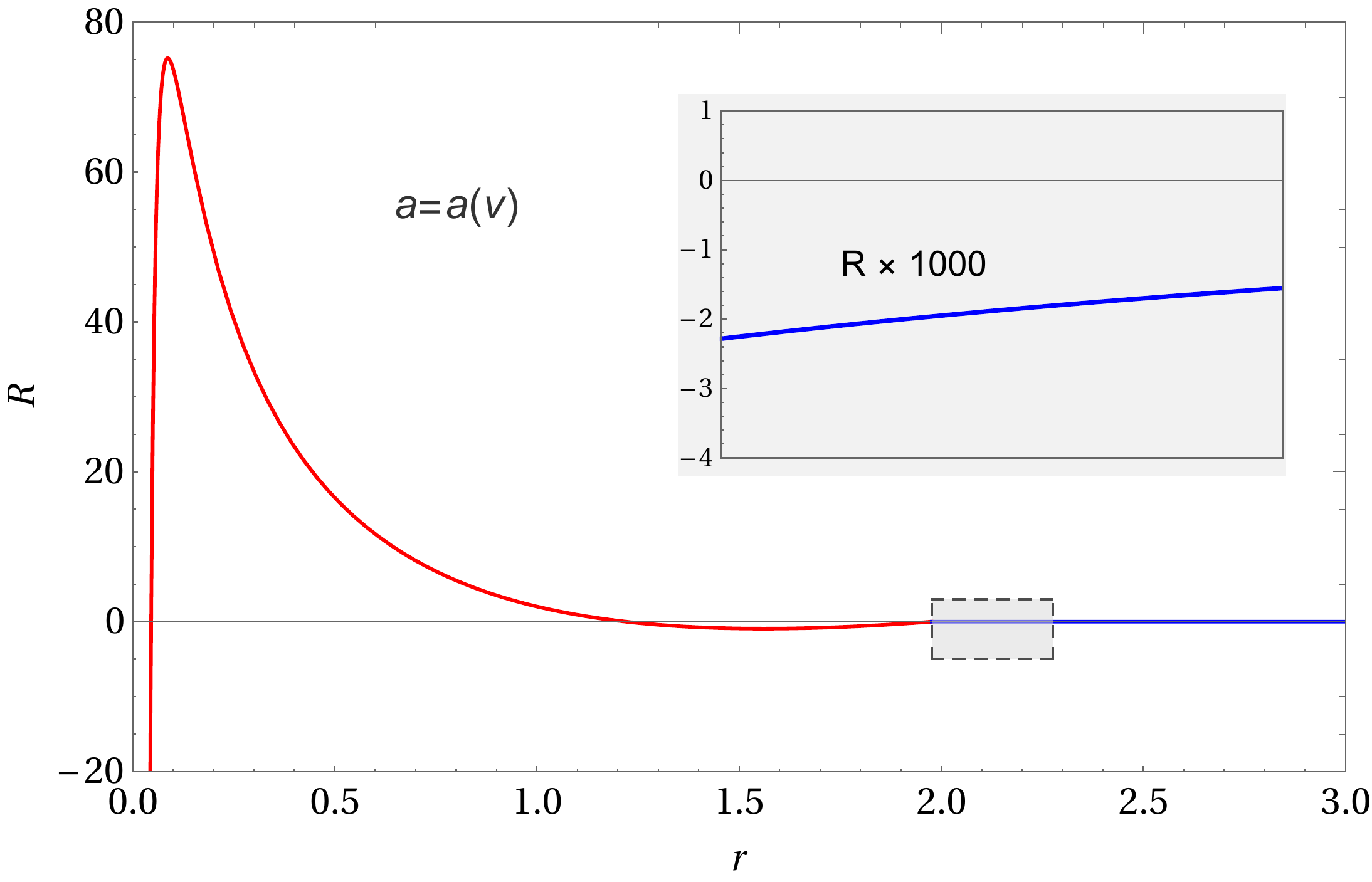} \
	\includegraphics[width=0.325\textwidth]{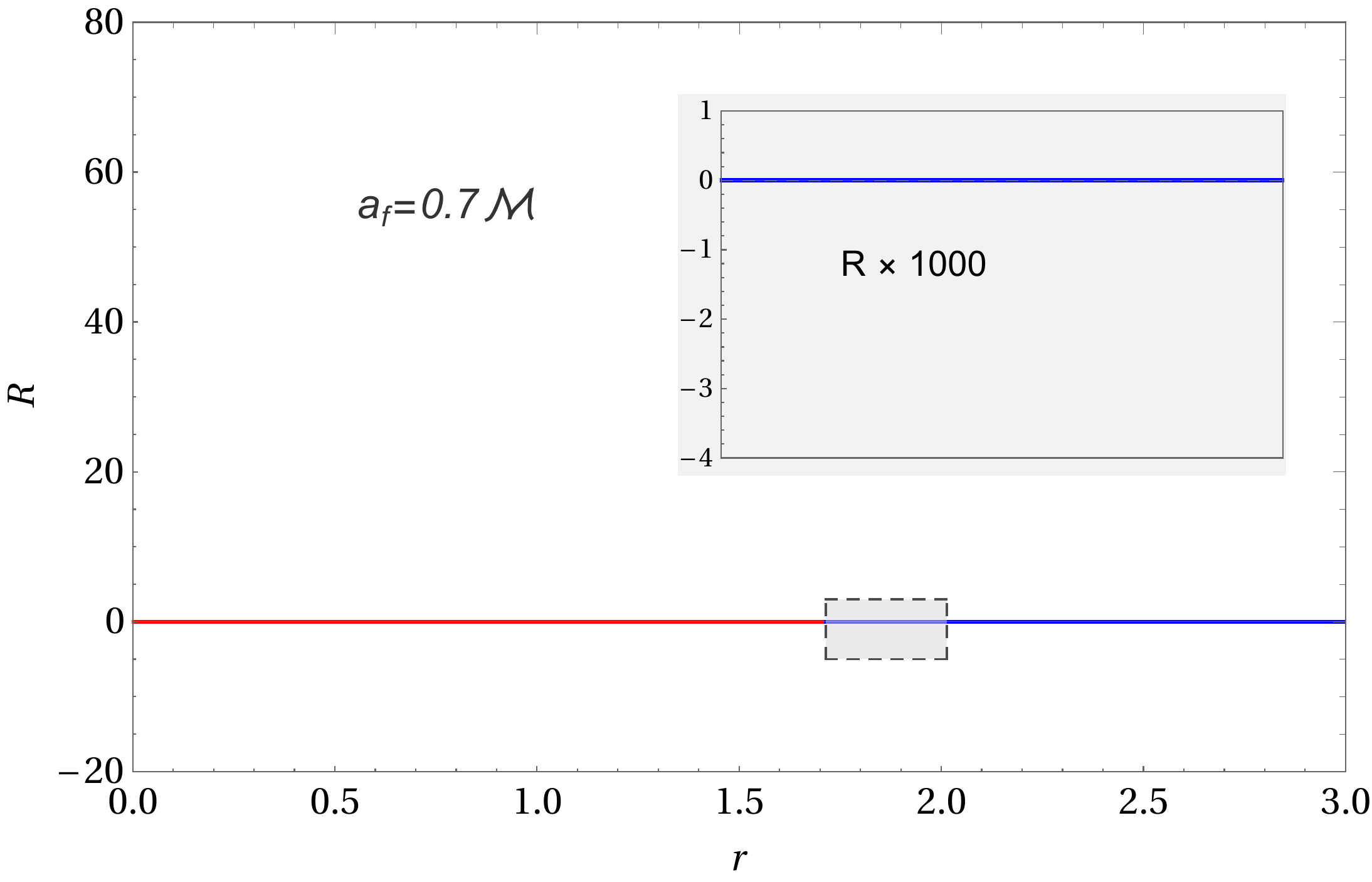} \
	\caption{\footnotesize Curvature $R$ in Eq.~\eqref{R-EF2} for $N=2$ (see Eq.~\eqref{m2} for the stationary case). The rotational parameter $a(v)$ and $n(v)$ are given by Eqs.~\eqref{a-Schw} and~\eqref{n-Kerr}, respectively, while $l(v)$ is taken as $l(v)=n(v)+1$, thereby satisfying the constraints~\eqref{order} and~\eqref{order2}. The spacetime evolves from an initial regular, spherically symmetric state (left panels) with $\alpha=3$, passes through a transient regime (central panel), and ultimately reaches the Kerr configuration with $a_f=0.7{\cal M}$. The upper plot shows a magnified view of the small boxed region in the corresponding lower plot, highlighting the exterior $r>h(v)$ (blue). The frequency is $\omega=0.1$, and the radial coordinate $r$ is measured in units of ${\cal M}$. }
	\label{fig4}
\end{figure*}
%
%
\section{Kerr BH formation} 
\label{sec4} 
\noindent  Since the collapse is characterized by $\dot{n}_i<0$, in addition to the singularities arising from the kinetic contribution in \eqref{R-EF2}, further singularities associated with the mass function  $m(v,r)$ develop whenever at least one of the functions $n_i(v)$ evolves through the interval $[-1,2]$ before reaching the Kerr configuration at $n_i=-1$. To investigate this process systematically, we introduce, in analogy with Eq.~\eqref{a(t)}, the function
\begin{equation} 
	\label{n2(t)} 
	n_i(v)=\frac{\alpha_i+\beta_i}{2}+\frac{\beta_i-\alpha_i}{2}\tanh(\omega{v})\ , \quad \forall\,i\ , 
\end{equation}
where $\alpha_i\equiv n_i(-\infty)$ and $\beta_i\equiv n_i(+\infty)$ characterize the initial and final values of $n_i(v)$, respectively, with at least one $\beta_i=-1$. Notice that the case discussed in Subsection~\ref{3-A} is recovered when $\alpha_i=\beta_i$ for all $i$.

Since the frequency $\omega$ appearing in Eqs.~\eqref{a(t)} and~\eqref{n2(t)} is the same, the quantities $\{a(v),n_i(v)\}$ are coupled and therefore Eq.~\eqref{n2(t)} can be expressed directly in terms of $a(v)$ as
\begin{equation} 
	\label{n3(t)} 
	n_i(v)=\left(\frac{\alpha_i{a_f}-\beta_i{a_0}}{a_f-a_0}\right)+\left(\frac{\beta_i-\alpha_i}{a_f-a_0}\right) a(v)\ , \quad \forall\,i\ . 
\end{equation}
It is important to emphasize that extending the discrete labels $n_i\in\mathbb{N}$ to the continuous functions
\begin{equation}
	n_i\in\mathbb{N}\Rightarrow n(v)=n(a(v))
\end{equation}
does not introduce any additional degrees of freedom or conserved charges. The collapse remains entirely characterized by the set $\{{\cal M},\,a(v)\}$, while the functions $n_i(v)$ are fully determined by the evolution of $a(v)$.

Owing to the hierarchy~\eqref{order}, the evolution is controlled by the leading element $n_1(v)\equiv n(v)$. By virtue of condition~\eqref{order2}, this parameter remains the smallest throughout the evolution and therefore controls the approach to the critical value $n=-1$, at which the Kerr BH emerges. Consequently, without loss of generality, we restrict the analysis to the minimal configuration $N=1$, for which the time-dependent extension of~\eqref{m1} takes the explicit form
\begin{equation}
	\label{m1(v)}
	m(v,r)=\frac{{\cal M}}{[n(v)-2]}\left[\frac{r^3}{h(v)^3}\left[n(v)+1\right]-3\left(\frac{r}{h(v)}\right)^{n(v)+1}\right],
\end{equation}
which yields
\begin{eqnarray}
	\label{Delta1}
	&&{\Delta}(v,r)=r^2+a^2\nonumber\\
	&&-\frac{2{\cal M}r}{[n(v)-2]}\left[\frac{r^3}{h(v)^3}\left[n(v)+1\right]-3\left(\frac{r}{h(v)}\right)^{n(v)+1}\right].
\end{eqnarray}
In particular, the formation of the Kerr BH from an initially regular Schwarzschild configuration is described by Eqs.~\eqref{a(t)} and~\eqref{n3(t)} with $a_0=0$ and $\beta=-1$, respectively, which reduce to
\begin{eqnarray}
	\label{a-Schw}
	&&a(v)=\frac{a_f}{2}\left[1+\tanh(\omega{v})\right]
	\\
	\label{n-Kerr}
	&&n(v)=\alpha-\frac{(1+\alpha)}{a_f} a(v)\ .
\end{eqnarray}
Although we focus here on a Schwarzschild initial state, the construction applies equally to rotating configurations with $a_0\neq0$, as follows directly from Eqs.~\eqref{a(t)} and~\eqref{n3(t)}. Expressions~\eqref{a-Schw} and~\eqref{n-Kerr} provide a detailed description of the evolution from a regular Schwarzschild BH to a Kerr BH, as illustrated in Figs.~\ref{fig3} and~\ref{fig4} for $N=1$ and $N=2$, respectively. The system must necessarily pass through a transient nonstationary singular stage, as revealed by the scalar curvature~\eqref{R-EF2}, before finally settling into the corresponding stationary Kerr configuration. The collapse is fully specified once a regular initial state, characterized by $\alpha\equiv n(-\infty)>2$, and a final rotational state specified by $a_f$ are prescribed. In particular, the extremal Kerr BH corresponds to $a_f={\cal M}$, in which case Eq.~\eqref{a-Schw} reduces to the expression in~\eqref{a(t)2}.

\section{Final remarks.} 

\noindent While we have developed an exact analytical model of axisymmetric gravitational collapse leading to the formation of the Kerr BH, including both the extremal case and the novel quasi-extremal regime with $h_c\sim\,h\not\approx{\cal M}$ [see Eqs.~\eqref{a(t)2},~\eqref{a(t)3},~\eqref{a-Schw}, and~\eqref{n-Kerr}], two key aspects remain to be addressed in order to complete this preliminary description. The first is to identify the source generating the geometry considered here and to assess how realistic it may be. The second is the rigorous determination of the causal structure. In this regard, our solution consistently exhibit two positive roots of $\Delta(v,r)=0$  throughout the collapse. The persistence of this structure strongly suggests that the curvature singularity remains confined within a trapped region closely associated with the evolving Kerr radius $h(v)={\cal M}+\sqrt{{\cal M}^2-a(v)^2}$. Establishing the existence of such a trapped surface would provide strong evidence that the singularity never becomes naked.

Finally, we emphasize that our geometric framework contains a curvature singularity associated with the kinematics of the collapse itself, as seen in Eq.~\eqref{R-EF2}, and one might therefore question its physical relevance. Our primary objective, however, is to investigate the formation of the Kerr BH, which necessarily entails the emergence of the singularity intrinsic to this solution. Since the curvature singularity has already formed while the spacetime is still evolving toward the Kerr BH, the present model should be interpreted not as a complete analytical description of gravitational collapse, which would include a prolonged regular stage, but rather as an exact analytical model of its final stage, immediately preceding the formation of the Kerr BH.

To conclude, we highlight two aspects explicitly identified through the curvature~\eqref{R-EF2} that deserve further investigation: (i) the kinematic contribution to the singularity at $r=0$, controlled by the term $R\sim{-2a\dot{a}/r^3}$, and (ii) the observational consequences of the transient exterior curvature $R\sim{1/r^2}$, given in Eq.~\eqref{Rfar}. The first calls for a rigorous analysis of the Raychaudhuri equation~\cite{Raychaudhuri:1953yv}, in particular of the contribution $\sim{R_{\mu\nu}u^\mu{u^\nu}}$. The second concerns how the curvature in Eq.~\eqref{Rfar} affects the propagation of particles and light, potentially giving rise to observable signatures associated with the formation of a rotating BH~\cite{Perlick:2004tq}.

\subsection*{Acknowledgments}
\vspace*{1mm}
This work was partially supported by ANID-FONDECYT Grant No. 1250227.
%

%
%
%
\bibliography{references.bib}

\begin{thebibliography}{47}%
\makeatletter
\providecommand \@ifxundefined [1]{%
 \@ifx{#1\undefined}
}%
\providecommand \@ifnum [1]{%
 \ifnum #1\expandafter \@firstoftwo
 \else \expandafter \@secondoftwo
 \fi
}%
\providecommand \@ifx [1]{%
 \ifx #1\expandafter \@firstoftwo
 \else \expandafter \@secondoftwo
 \fi
}%
\providecommand \natexlab [1]{#1}%
\providecommand \enquote  [1]{``#1''}%
\providecommand \bibnamefont  [1]{#1}%
\providecommand \bibfnamefont [1]{#1}%
\providecommand \citenamefont [1]{#1}%
\providecommand \href@noop [0]{\@secondoftwo}%
\providecommand \href [0]{\begingroup \@sanitize@url \@href}%
\providecommand \@href[1]{\@@startlink{#1}\@@href}%
\providecommand \@@href[1]{\endgroup#1\@@endlink}%
\providecommand \@sanitize@url [0]{\catcode `\\12\catcode `\$12\catcode
  `\&12\catcode `\#12\catcode `\^12\catcode `\_12\catcode `\%12\relax}%
\providecommand \@@startlink[1]{}%
\providecommand \@@endlink[0]{}%
\providecommand \url  [0]{\begingroup\@sanitize@url \@url }%
\providecommand \@url [1]{\endgroup\@href {#1}{\urlprefix }}%
\providecommand \urlprefix  [0]{URL }%
\providecommand \Eprint [0]{\href }%
\providecommand \doibase [0]{http://dx.doi.org/}%
\providecommand \selectlanguage [0]{\@gobble}%
\providecommand \bibinfo  [0]{\@secondoftwo}%
\providecommand \bibfield  [0]{\@secondoftwo}%
\providecommand \translation [1]{[#1]}%
\providecommand \BibitemOpen [0]{}%
\providecommand \bibitemStop [0]{}%
\providecommand \bibitemNoStop [0]{.\EOS\space}%
\providecommand \EOS [0]{\spacefactor3000\relax}%
\providecommand \BibitemShut  [1]{\csname bibitem#1\endcsname}%
\let\auto@bib@innerbib\@empty
\bibitem [{\citenamefont {Israel}(1967)}]{Israel:1967wq}%
  \BibitemOpen
  \bibfield  {author} {\bibinfo {author} {\bibfnamefont {W.}~\bibnamefont
  {Israel}},\ }\href {\doibase 10.1103/PhysRev.164.1776} {\bibfield  {journal}
  {\bibinfo  {journal} {Phys. Rev.}\ }\textbf {\bibinfo {volume} {164}},\
  \bibinfo {pages} {1776} (\bibinfo {year} {1967})}\BibitemShut {NoStop}%
\bibitem [{\citenamefont {Israel}(1968)}]{Israel:1967za}%
  \BibitemOpen
  \bibfield  {author} {\bibinfo {author} {\bibfnamefont {W.}~\bibnamefont
  {Israel}},\ }\href {\doibase 10.1007/BF01645859} {\bibfield  {journal}
  {\bibinfo  {journal} {Commun. Math. Phys.}\ }\textbf {\bibinfo {volume}
  {8}},\ \bibinfo {pages} {245} (\bibinfo {year} {1968})}\BibitemShut {NoStop}%
\bibitem [{\citenamefont {Carter}(1971)}]{Carter:1971zc}%
  \BibitemOpen
  \bibfield  {author} {\bibinfo {author} {\bibfnamefont {B.}~\bibnamefont
  {Carter}},\ }\href {\doibase 10.1103/PhysRevLett.26.331} {\bibfield
  {journal} {\bibinfo  {journal} {Phys. Rev. Lett.}\ }\textbf {\bibinfo
  {volume} {26}},\ \bibinfo {pages} {331} (\bibinfo {year} {1971})}\BibitemShut
  {NoStop}%
\bibitem [{\citenamefont {Hawking}(1972)}]{Hawking:1971vc}%
  \BibitemOpen
  \bibfield  {author} {\bibinfo {author} {\bibfnamefont {S.}~\bibnamefont
  {Hawking}},\ }\href {\doibase 10.1007/BF01877517} {\bibfield  {journal}
  {\bibinfo  {journal} {Commun. Math. Phys.}\ }\textbf {\bibinfo {volume}
  {25}},\ \bibinfo {pages} {152} (\bibinfo {year} {1972})}\BibitemShut
  {NoStop}%
\bibitem [{\citenamefont {Robinson}(1975)}]{Robinson:1975bv}%
  \BibitemOpen
  \bibfield  {author} {\bibinfo {author} {\bibfnamefont {D.~C.}\ \bibnamefont
  {Robinson}},\ }\href {\doibase 10.1103/PhysRevLett.34.905} {\bibfield
  {journal} {\bibinfo  {journal} {Phys. Rev. Lett.}\ }\textbf {\bibinfo
  {volume} {34}},\ \bibinfo {pages} {905} (\bibinfo {year} {1975})}\BibitemShut
  {NoStop}%
\bibitem [{\citenamefont {Heusler}(1996)}]{Heusler:1996jaf}%
  \BibitemOpen
  \bibfield  {author} {\bibinfo {author} {\bibfnamefont {M.}~\bibnamefont
  {Heusler}},\ }\href {\doibase 10.1017/cbo9780511661396} {\emph {\bibinfo
  {title} {{Black Hole Uniqueness Theorems}}}}\ (\bibinfo {year}
  {1996})\BibitemShut {NoStop}%
\bibitem [{\citenamefont {Chrusciel}\ \emph {et~al.}(2012)\citenamefont
  {Chrusciel}, \citenamefont {Lopes~Costa},\ and\ \citenamefont
  {Heusler}}]{Chrusciel:2012jk}%
  \BibitemOpen
  \bibfield  {author} {\bibinfo {author} {\bibfnamefont {P.~T.}\ \bibnamefont
  {Chrusciel}}, \bibinfo {author} {\bibfnamefont {J.}~\bibnamefont
  {Lopes~Costa}}, \ and\ \bibinfo {author} {\bibfnamefont {M.}~\bibnamefont
  {Heusler}},\ }\href {\doibase 10.12942/lrr-2012-7} {\bibfield  {journal}
  {\bibinfo  {journal} {Living Rev. Rel.}\ }\textbf {\bibinfo {volume} {15}},\
  \bibinfo {pages} {7} (\bibinfo {year} {2012})},\ \Eprint
  {http://arxiv.org/abs/1205.6112} {arXiv:1205.6112 [gr-qc]} \BibitemShut
  {NoStop}%
\bibitem [{\citenamefont {Penrose}(1965)}]{Penrose:1964wq}%
  \BibitemOpen
  \bibfield  {author} {\bibinfo {author} {\bibfnamefont {R.}~\bibnamefont
  {Penrose}},\ }\href {\doibase 10.1103/PhysRevLett.14.57} {\bibfield
  {journal} {\bibinfo  {journal} {Phys. Rev. Lett.}\ }\textbf {\bibinfo
  {volume} {14}},\ \bibinfo {pages} {57} (\bibinfo {year} {1965})}\BibitemShut
  {NoStop}%
\bibitem [{\citenamefont {Hawking}\ and\ \citenamefont
  {Penrose}(1970)}]{Hawking:1970zqf}%
  \BibitemOpen
  \bibfield  {author} {\bibinfo {author} {\bibfnamefont {S.~W.}\ \bibnamefont
  {Hawking}}\ and\ \bibinfo {author} {\bibfnamefont {R.}~\bibnamefont
  {Penrose}},\ }\href {\doibase 10.1098/rspa.1970.0021} {\bibfield  {journal}
  {\bibinfo  {journal} {Proc. Roy. Soc. Lond. A}\ }\textbf {\bibinfo {volume}
  {314}},\ \bibinfo {pages} {529} (\bibinfo {year} {1970})}\BibitemShut
  {NoStop}%
\bibitem [{\citenamefont {Hawking}\ and\ \citenamefont
  {Ellis}(2011)}]{Hawking:1973uf}%
  \BibitemOpen
  \bibfield  {author} {\bibinfo {author} {\bibfnamefont {S.~W.}\ \bibnamefont
  {Hawking}}\ and\ \bibinfo {author} {\bibfnamefont {G.~F.~R.}\ \bibnamefont
  {Ellis}},\ }\href {\doibase 10.1017/CBO9780511524646} {\emph {\bibinfo
  {title} {{The Large Scale Structure of Space-Time}}}},\ Cambridge Monographs
  on Mathematical Physics\ (\bibinfo  {publisher} {Cambridge University
  Press},\ \bibinfo {year} {2011})\BibitemShut {NoStop}%
\bibitem [{\citenamefont {Penrose}(1969)}]{Penrose:1969pc}%
  \BibitemOpen
  \bibfield  {author} {\bibinfo {author} {\bibfnamefont {R.}~\bibnamefont
  {Penrose}},\ }\href {\doibase 10.1023/A:1016578408204} {\bibfield  {journal}
  {\bibinfo  {journal} {Riv. Nuovo Cim.}\ }\textbf {\bibinfo {volume} {1}},\
  \bibinfo {pages} {252} (\bibinfo {year} {1969})}\BibitemShut {NoStop}%
\bibitem [{\citenamefont {Kerr}(1963)}]{Kerr:1963ud}%
  \BibitemOpen
  \bibfield  {author} {\bibinfo {author} {\bibfnamefont {R.~P.}\ \bibnamefont
  {Kerr}},\ }\href {\doibase 10.1103/PhysRevLett.11.237} {\bibfield  {journal}
  {\bibinfo  {journal} {Phys. Rev. Lett.}\ }\textbf {\bibinfo {volume} {11}},\
  \bibinfo {pages} {237} (\bibinfo {year} {1963})}\BibitemShut {NoStop}%
\bibitem [{\citenamefont {Oppenheimer}\ and\ \citenamefont
  {Snyder}(1939)}]{Oppenheimer:1939ue}%
  \BibitemOpen
  \bibfield  {author} {\bibinfo {author} {\bibfnamefont {J.~R.}\ \bibnamefont
  {Oppenheimer}}\ and\ \bibinfo {author} {\bibfnamefont {H.}~\bibnamefont
  {Snyder}},\ }\href {\doibase 10.1103/PhysRev.56.455} {\bibfield  {journal}
  {\bibinfo  {journal} {Phys. Rev.}\ }\textbf {\bibinfo {volume} {56}},\
  \bibinfo {pages} {455} (\bibinfo {year} {1939})}\BibitemShut {NoStop}%
\bibitem [{\citenamefont {Christodoulou}(1984)}]{Christodoulou:1984mz}%
  \BibitemOpen
  \bibfield  {author} {\bibinfo {author} {\bibfnamefont {D.}~\bibnamefont
  {Christodoulou}},\ }\href {\doibase 10.1007/BF01223743} {\bibfield  {journal}
  {\bibinfo  {journal} {Commun. Math. Phys.}\ }\textbf {\bibinfo {volume}
  {93}},\ \bibinfo {pages} {171} (\bibinfo {year} {1984})}\BibitemShut
  {NoStop}%
\bibitem [{\citenamefont {Joshi}\ and\ \citenamefont
  {Dwivedi}(1993)}]{Joshi:1993zg}%
  \BibitemOpen
  \bibfield  {author} {\bibinfo {author} {\bibfnamefont {P.~S.}\ \bibnamefont
  {Joshi}}\ and\ \bibinfo {author} {\bibfnamefont {I.~H.}\ \bibnamefont
  {Dwivedi}},\ }\href {\doibase 10.1103/PhysRevD.47.5357} {\bibfield  {journal}
  {\bibinfo  {journal} {Phys. Rev. D}\ }\textbf {\bibinfo {volume} {47}},\
  \bibinfo {pages} {5357} (\bibinfo {year} {1993})},\ \Eprint
  {http://arxiv.org/abs/gr-qc/9303037} {arXiv:gr-qc/9303037} \BibitemShut
  {NoStop}%
\bibitem [{\citenamefont {Joshi}\ \emph {et~al.}(2002)\citenamefont {Joshi},
  \citenamefont {Dadhich},\ and\ \citenamefont {Maartens}}]{Joshi:2001xi}%
  \BibitemOpen
  \bibfield  {author} {\bibinfo {author} {\bibfnamefont {P.~S.}\ \bibnamefont
  {Joshi}}, \bibinfo {author} {\bibfnamefont {N.}~\bibnamefont {Dadhich}}, \
  and\ \bibinfo {author} {\bibfnamefont {R.}~\bibnamefont {Maartens}},\ }\href
  {\doibase 10.1103/PhysRevD.65.101501} {\bibfield  {journal} {\bibinfo
  {journal} {Phys. Rev. D}\ }\textbf {\bibinfo {volume} {65}},\ \bibinfo
  {pages} {101501} (\bibinfo {year} {2002})},\ \Eprint
  {http://arxiv.org/abs/gr-qc/0109051} {arXiv:gr-qc/0109051} \BibitemShut
  {NoStop}%
\bibitem [{\citenamefont {Mena}\ \emph {et~al.}(2004)\citenamefont {Mena},
  \citenamefont {Nolan},\ and\ \citenamefont {Tavakol}}]{Mena:2004ck}%
  \BibitemOpen
  \bibfield  {author} {\bibinfo {author} {\bibfnamefont {F.~C.}\ \bibnamefont
  {Mena}}, \bibinfo {author} {\bibfnamefont {B.~C.}\ \bibnamefont {Nolan}}, \
  and\ \bibinfo {author} {\bibfnamefont {R.}~\bibnamefont {Tavakol}},\ }\href
  {\doibase 10.1103/PhysRevD.70.084030} {\bibfield  {journal} {\bibinfo
  {journal} {Phys. Rev. D}\ }\textbf {\bibinfo {volume} {70}},\ \bibinfo
  {pages} {084030} (\bibinfo {year} {2004})},\ \Eprint
  {http://arxiv.org/abs/gr-qc/0405041} {arXiv:gr-qc/0405041} \BibitemShut
  {NoStop}%
\bibitem [{\citenamefont {Lasky}\ and\ \citenamefont
  {Lun}(2006)}]{Lasky:2006mg}%
  \BibitemOpen
  \bibfield  {author} {\bibinfo {author} {\bibfnamefont {P.~D.}\ \bibnamefont
  {Lasky}}\ and\ \bibinfo {author} {\bibfnamefont {A.~W.~C.}\ \bibnamefont
  {Lun}},\ }\href {\doibase 10.1103/PhysRevD.74.084013} {\bibfield  {journal}
  {\bibinfo  {journal} {Phys. Rev. D}\ }\textbf {\bibinfo {volume} {74}},\
  \bibinfo {pages} {084013} (\bibinfo {year} {2006})},\ \Eprint
  {http://arxiv.org/abs/gr-qc/0606055} {arXiv:gr-qc/0606055} \BibitemShut
  {NoStop}%
\bibitem [{\citenamefont {Joshi}(2012)}]{Joshi:2008zz}%
  \BibitemOpen
  \bibinfo {editor} {\bibfnamefont {P.~S.}\ \bibnamefont {Joshi}},\ ed.,\ \href
  {\doibase 10.1017/CBO9780511536274} {\emph {\bibinfo {title} {{Gravitational
  Collapse and Spacetime Singularities}}}},\ Cambridge Monographs on
  Mathematical Physics\ (\bibinfo  {publisher} {Cambridge University Press},\
  \bibinfo {year} {2012})\BibitemShut {NoStop}%
\bibitem [{\citenamefont {Mosani}\ \emph {et~al.}(2020)\citenamefont {Mosani},
  \citenamefont {Dey},\ and\ \citenamefont {Joshi}}]{Mosani:2020ena}%
  \BibitemOpen
  \bibfield  {author} {\bibinfo {author} {\bibfnamefont {K.}~\bibnamefont
  {Mosani}}, \bibinfo {author} {\bibfnamefont {D.}~\bibnamefont {Dey}}, \ and\
  \bibinfo {author} {\bibfnamefont {P.~S.}\ \bibnamefont {Joshi}},\ }\href
  {\doibase 10.1103/PhysRevD.101.044052} {\bibfield  {journal} {\bibinfo
  {journal} {Phys. Rev. D}\ }\textbf {\bibinfo {volume} {101}},\ \bibinfo
  {pages} {044052} (\bibinfo {year} {2020})},\ \bibinfo {note} {[Erratum:
  Phys.Rev.D 107, 069903 (2023)]},\ \Eprint {http://arxiv.org/abs/2001.04367}
  {arXiv:2001.04367 [gr-qc]} \BibitemShut {NoStop}%
\bibitem [{\citenamefont {Joshi}\ \emph {et~al.}(2024)\citenamefont {Joshi},
  \citenamefont {Mosani},\ and\ \citenamefont {Joshi}}]{Joshi:2023ugm}%
  \BibitemOpen
  \bibfield  {author} {\bibinfo {author} {\bibfnamefont {A.~B.}\ \bibnamefont
  {Joshi}}, \bibinfo {author} {\bibfnamefont {K.}~\bibnamefont {Mosani}}, \
  and\ \bibinfo {author} {\bibfnamefont {P.~S.}\ \bibnamefont {Joshi}},\ }\href
  {\doibase 10.1103/PhysRevD.109.064019} {\bibfield  {journal} {\bibinfo
  {journal} {Phys. Rev. D}\ }\textbf {\bibinfo {volume} {109}},\ \bibinfo
  {pages} {064019} (\bibinfo {year} {2024})},\ \Eprint
  {http://arxiv.org/abs/2310.01222} {arXiv:2310.01222 [gr-qc]} \BibitemShut
  {NoStop}%
\bibitem [{\citenamefont {Teukolsky}(2015)}]{Teukolsky:2014vca}%
  \BibitemOpen
  \bibfield  {author} {\bibinfo {author} {\bibfnamefont {S.~A.}\ \bibnamefont
  {Teukolsky}},\ }\href {\doibase 10.1088/0264-9381/32/12/124006} {\bibfield
  {journal} {\bibinfo  {journal} {Class. Quant. Grav.}\ }\textbf {\bibinfo
  {volume} {32}},\ \bibinfo {pages} {124006} (\bibinfo {year} {2015})},\
  \Eprint {http://arxiv.org/abs/1410.2130} {arXiv:1410.2130 [gr-qc]}
  \BibitemShut {NoStop}%
\bibitem [{\citenamefont {Ovalle}(2024)}]{Ovalle:2024wtv}%
  \BibitemOpen
  \bibfield  {author} {\bibinfo {author} {\bibfnamefont {J.}~\bibnamefont
  {Ovalle}},\ }\href {\doibase 10.1103/PhysRevD.109.104032} {\bibfield
  {journal} {\bibinfo  {journal} {Phys. Rev. D}\ }\textbf {\bibinfo {volume}
  {109}},\ \bibinfo {pages} {104032} (\bibinfo {year} {2024})}\BibitemShut
  {NoStop}%
\bibitem [{\citenamefont {Ovalle}(2025)}]{Ovalle:2025pue}%
  \BibitemOpen
  \bibfield  {author} {\bibinfo {author} {\bibfnamefont {J.}~\bibnamefont
  {Ovalle}},\ }\href@noop {} {\  (\bibinfo {year} {2025})},\ \Eprint
  {http://arxiv.org/abs/2509.00816} {arXiv:2509.00816 [gr-qc]} \BibitemShut
  {NoStop}%
\bibitem [{\citenamefont {Ovalle}\ \emph {et~al.}(2026)\citenamefont {Ovalle},
  \citenamefont {Casadio},\ and\ \citenamefont {Kamenshchik}}]{Ovalle:2026lxb}%
  \BibitemOpen
  \bibfield  {author} {\bibinfo {author} {\bibfnamefont {J.}~\bibnamefont
  {Ovalle}}, \bibinfo {author} {\bibfnamefont {R.}~\bibnamefont {Casadio}}, \
  and\ \bibinfo {author} {\bibfnamefont {A.}~\bibnamefont {Kamenshchik}},\
  }\href {\doibase 10.1103/cbs6-d7pr} {\bibfield  {journal} {\bibinfo
  {journal} {Phys. Rev. D}\ }\textbf {\bibinfo {volume} {113}},\ \bibinfo
  {pages} {064042} (\bibinfo {year} {2026})},\ \Eprint
  {http://arxiv.org/abs/2603.06451} {arXiv:2603.06451 [gr-qc]} \BibitemShut
  {NoStop}%
\bibitem [{\citenamefont {Casadio}\ \emph {et~al.}(2026)\citenamefont
  {Casadio}, \citenamefont {Giusti}, \citenamefont {Kamenshchik},\ and\
  \citenamefont {Ovalle}}]{Casadio:2026tmd}%
  \BibitemOpen
  \bibfield  {author} {\bibinfo {author} {\bibfnamefont {R.}~\bibnamefont
  {Casadio}}, \bibinfo {author} {\bibfnamefont {A.}~\bibnamefont {Giusti}},
  \bibinfo {author} {\bibfnamefont {A.}~\bibnamefont {Kamenshchik}}, \ and\
  \bibinfo {author} {\bibfnamefont {J.}~\bibnamefont {Ovalle}},\ }\href@noop {}
  {\  (\bibinfo {year} {2026})},\ \Eprint {http://arxiv.org/abs/2605.01808}
  {arXiv:2605.01808 [gr-qc]} \BibitemShut {NoStop}%
\bibitem [{\citenamefont {Lobo}\ and\ \citenamefont
  {Rodrigues}(2026{\natexlab{a}})}]{Lobo:2026dnl}%
  \BibitemOpen
  \bibfield  {author} {\bibinfo {author} {\bibfnamefont {F.~S.~N.}\
  \bibnamefont {Lobo}}\ and\ \bibinfo {author} {\bibfnamefont {M.~E.}\
  \bibnamefont {Rodrigues}},\ }\href@noop {} {\  (\bibinfo {year}
  {2026}{\natexlab{a}})},\ \Eprint {http://arxiv.org/abs/2607.14079}
  {arXiv:2607.14079 [gr-qc]} \BibitemShut {NoStop}%
\bibitem [{\citenamefont {Lobo}\ and\ \citenamefont
  {Rodrigues}(2026{\natexlab{b}})}]{Lobo:2026iuy}%
  \BibitemOpen
  \bibfield  {author} {\bibinfo {author} {\bibfnamefont {F.~S.~N.}\
  \bibnamefont {Lobo}}\ and\ \bibinfo {author} {\bibfnamefont {M.~E.}\
  \bibnamefont {Rodrigues}},\ }\href@noop {} {\  (\bibinfo {year}
  {2026}{\natexlab{b}})},\ \Eprint {http://arxiv.org/abs/2607.24349}
  {arXiv:2607.24349 [gr-qc]} \BibitemShut {NoStop}%
\bibitem [{\citenamefont {Ovalle}(2026)}]{Ovalle:2026ajv}%
  \BibitemOpen
  \bibfield  {author} {\bibinfo {author} {\bibfnamefont {J.}~\bibnamefont
  {Ovalle}},\ }\href@noop {} {\  (\bibinfo {year} {2026})},\ \Eprint
  {http://arxiv.org/abs/2607.14910} {arXiv:2607.14910 [gr-qc]} \BibitemShut
  {NoStop}%
\bibitem [{\citenamefont {Gurses}\ and\ \citenamefont
  {Gursey}(1975)}]{Gurses:1975vu}%
  \BibitemOpen
  \bibfield  {author} {\bibinfo {author} {\bibfnamefont {M.}~\bibnamefont
  {Gurses}}\ and\ \bibinfo {author} {\bibfnamefont {F.}~\bibnamefont
  {Gursey}},\ }\href {\doibase 10.1063/1.522480} {\bibfield  {journal}
  {\bibinfo  {journal} {J. Math. Phys.}\ }\textbf {\bibinfo {volume} {16}},\
  \bibinfo {pages} {2385} (\bibinfo {year} {1975})}\BibitemShut {NoStop}%
\bibitem [{\citenamefont {Casadio}\ \emph {et~al.}(2024)\citenamefont
  {Casadio}, \citenamefont {Kamenshchik},\ and\ \citenamefont
  {Ovalle}}]{Casadio:2024fol}%
  \BibitemOpen
  \bibfield  {author} {\bibinfo {author} {\bibfnamefont {R.}~\bibnamefont
  {Casadio}}, \bibinfo {author} {\bibfnamefont {A.}~\bibnamefont
  {Kamenshchik}}, \ and\ \bibinfo {author} {\bibfnamefont {J.}~\bibnamefont
  {Ovalle}},\ }\href {\doibase 10.1103/PhysRevD.110.044001} {\bibfield
  {journal} {\bibinfo  {journal} {Phys. Rev. D}\ }\textbf {\bibinfo {volume}
  {110}},\ \bibinfo {pages} {044001} (\bibinfo {year} {2024})}\BibitemShut
  {NoStop}%
\bibitem [{\citenamefont {Casadio}\ \emph {et~al.}(2025)\citenamefont
  {Casadio}, \citenamefont {Kamenshchik},\ and\ \citenamefont
  {Ovalle}}]{Casadio:2025pun}%
  \BibitemOpen
  \bibfield  {author} {\bibinfo {author} {\bibfnamefont {R.}~\bibnamefont
  {Casadio}}, \bibinfo {author} {\bibfnamefont {A.}~\bibnamefont
  {Kamenshchik}}, \ and\ \bibinfo {author} {\bibfnamefont {J.}~\bibnamefont
  {Ovalle}},\ }\href {\doibase 10.1103/PhysRevD.111.064036} {\bibfield
  {journal} {\bibinfo  {journal} {Phys. Rev. D}\ }\textbf {\bibinfo {volume}
  {111}},\ \bibinfo {pages} {064036} (\bibinfo {year} {2025})}\BibitemShut
  {NoStop}%
\bibitem [{\citenamefont {Lukash}\ and\ \citenamefont
  {Strokov}(2013)}]{Lukash:2013ts}%
  \BibitemOpen
  \bibfield  {author} {\bibinfo {author} {\bibfnamefont {V.~N.}\ \bibnamefont
  {Lukash}}\ and\ \bibinfo {author} {\bibfnamefont {V.~N.}\ \bibnamefont
  {Strokov}},\ }\href {\doibase 10.1142/S0217751X13500073} {\bibfield
  {journal} {\bibinfo  {journal} {Int. J. Mod. Phys. A}\ }\textbf {\bibinfo
  {volume} {28}},\ \bibinfo {pages} {1350007} (\bibinfo {year} {2013})},\
  \Eprint {http://arxiv.org/abs/1301.5544} {arXiv:1301.5544 [gr-qc]}
  \BibitemShut {NoStop}%
\bibitem [{\citenamefont {Ovalle}(2023)}]{Ovalle:2023vvu}%
  \BibitemOpen
  \bibfield  {author} {\bibinfo {author} {\bibfnamefont {J.}~\bibnamefont
  {Ovalle}},\ }\href {\doibase 10.1103/PhysRevD.107.104005} {\bibfield
  {journal} {\bibinfo  {journal} {Phys. Rev. D}\ }\textbf {\bibinfo {volume}
  {107}},\ \bibinfo {pages} {104005} (\bibinfo {year} {2023})}\BibitemShut
  {NoStop}%
\bibitem [{\citenamefont {Arrechea}\ \emph {et~al.}(2025)\citenamefont
  {Arrechea}, \citenamefont {Liberati}, \citenamefont {Neshat},\ and\
  \citenamefont {Vellucci}}]{Arrechea:2025fkk}%
  \BibitemOpen
  \bibfield  {author} {\bibinfo {author} {\bibfnamefont {J.}~\bibnamefont
  {Arrechea}}, \bibinfo {author} {\bibfnamefont {S.}~\bibnamefont {Liberati}},
  \bibinfo {author} {\bibfnamefont {H.}~\bibnamefont {Neshat}}, \ and\ \bibinfo
  {author} {\bibfnamefont {V.}~\bibnamefont {Vellucci}},\ }\href {\doibase
  10.1103/wk7j-yg1t} {\bibfield  {journal} {\bibinfo  {journal} {Phys. Rev. D}\
  }\textbf {\bibinfo {volume} {112}},\ \bibinfo {pages} {044024} (\bibinfo
  {year} {2025})}\BibitemShut {NoStop}%
\bibitem [{\citenamefont {Chandrasekhar}(1983)}]{Chandrasekhar:1983}%
  \BibitemOpen
  \bibfield  {author} {\bibinfo {author} {\bibfnamefont {S.}~\bibnamefont
  {Chandrasekhar}},\ }\href@noop {} {\emph {\bibinfo {title} {The Mathematical
  Theory of Black Holes}}},\ \bibinfo {series} {International Series of
  Monographs on Physics}, Vol.~\bibinfo {volume} {69}\ (\bibinfo  {publisher}
  {Oxford University Press},\ \bibinfo {address} {New York},\ \bibinfo {year}
  {1983})\BibitemShut {NoStop}%
\bibitem [{\citenamefont {Stephani}\ \emph {et~al.}(2003)\citenamefont
  {Stephani}, \citenamefont {Kramer}, \citenamefont {MacCallum}, \citenamefont
  {Hoenselaers},\ and\ \citenamefont {Herlt}}]{Stephani:2003tm}%
  \BibitemOpen
  \bibfield  {author} {\bibinfo {author} {\bibfnamefont {H.}~\bibnamefont
  {Stephani}}, \bibinfo {author} {\bibfnamefont {D.}~\bibnamefont {Kramer}},
  \bibinfo {author} {\bibfnamefont {M.~A.~H.}\ \bibnamefont {MacCallum}},
  \bibinfo {author} {\bibfnamefont {C.}~\bibnamefont {Hoenselaers}}, \ and\
  \bibinfo {author} {\bibfnamefont {E.}~\bibnamefont {Herlt}},\ }\href
  {\doibase 10.1017/CBO9780511535185} {\emph {\bibinfo {title} {{Exact
  solutions of Einstein's field equations}}}},\ Cambridge Monographs on
  Mathematical Physics\ (\bibinfo  {publisher} {Cambridge Univ. Press},\
  \bibinfo {address} {Cambridge},\ \bibinfo {year} {2003})\BibitemShut
  {NoStop}%
\bibitem [{\citenamefont {Bardeen}\ \emph {et~al.}(1973)\citenamefont
  {Bardeen}, \citenamefont {Carter},\ and\ \citenamefont
  {Hawking}}]{Bardeen:1973gs}%
  \BibitemOpen
  \bibfield  {author} {\bibinfo {author} {\bibfnamefont {J.~M.}\ \bibnamefont
  {Bardeen}}, \bibinfo {author} {\bibfnamefont {B.}~\bibnamefont {Carter}}, \
  and\ \bibinfo {author} {\bibfnamefont {S.~W.}\ \bibnamefont {Hawking}},\
  }\href {\doibase 10.1007/BF01645742} {\bibfield  {journal} {\bibinfo
  {journal} {Commun. Math. Phys.}\ }\textbf {\bibinfo {volume} {31}},\ \bibinfo
  {pages} {161} (\bibinfo {year} {1973})}\BibitemShut {NoStop}%
\bibitem [{\citenamefont {Hayward}(1994)}]{Hayward:1993wb}%
  \BibitemOpen
  \bibfield  {author} {\bibinfo {author} {\bibfnamefont {S.~A.}\ \bibnamefont
  {Hayward}},\ }\href {\doibase 10.1103/PhysRevD.49.6467} {\bibfield  {journal}
  {\bibinfo  {journal} {Phys. Rev. D}\ }\textbf {\bibinfo {volume} {49}},\
  \bibinfo {pages} {6467} (\bibinfo {year} {1994})}\BibitemShut {NoStop}%
\bibitem [{\citenamefont {Ashtekar}\ and\ \citenamefont
  {Krishnan}(2003)}]{Ashtekar:2003hk}%
  \BibitemOpen
  \bibfield  {author} {\bibinfo {author} {\bibfnamefont {A.}~\bibnamefont
  {Ashtekar}}\ and\ \bibinfo {author} {\bibfnamefont {B.}~\bibnamefont
  {Krishnan}},\ }\href {\doibase 10.1103/PhysRevD.68.104030} {\bibfield
  {journal} {\bibinfo  {journal} {Phys. Rev. D}\ }\textbf {\bibinfo {volume}
  {68}},\ \bibinfo {pages} {104030} (\bibinfo {year} {2003})},\ \Eprint
  {http://arxiv.org/abs/gr-qc/0308033} {arXiv:gr-qc/0308033} \BibitemShut
  {NoStop}%
\bibitem [{\citenamefont {Senovilla}\ and\ \citenamefont
  {Torres}(2015)}]{Senovilla:2014ika}%
  \BibitemOpen
  \bibfield  {author} {\bibinfo {author} {\bibfnamefont {J.~M.~M.}\
  \bibnamefont {Senovilla}}\ and\ \bibinfo {author} {\bibfnamefont
  {R.}~\bibnamefont {Torres}},\ }\href {\doibase 10.1088/0264-9381/32/8/085004}
  {\bibfield  {journal} {\bibinfo  {journal} {Class. Quant. Grav.}\ }\textbf
  {\bibinfo {volume} {32}},\ \bibinfo {pages} {085004} (\bibinfo {year}
  {2015})},\ \bibinfo {note} {[Erratum: Class.Quant.Grav. 32, 189501 (2015)]},\
  \Eprint {http://arxiv.org/abs/1409.6044} {arXiv:1409.6044 [gr-qc]}
  \BibitemShut {NoStop}%
\bibitem [{\citenamefont {Ben~Achour}\ \emph {et~al.}(2026)\citenamefont
  {Ben~Achour}, \citenamefont {Cisterna},\ and\ \citenamefont
  {Hassaine}}]{BenAchour:2025vur}%
  \BibitemOpen
  \bibfield  {author} {\bibinfo {author} {\bibfnamefont {J.}~\bibnamefont
  {Ben~Achour}}, \bibinfo {author} {\bibfnamefont {A.}~\bibnamefont
  {Cisterna}}, \ and\ \bibinfo {author} {\bibfnamefont {M.}~\bibnamefont
  {Hassaine}},\ }\href {\doibase 10.1088/1475-7516/2026/06/061} {\bibfield
  {journal} {\bibinfo  {journal} {JCAP}\ }\textbf {\bibinfo {volume} {06}},\
  \bibinfo {pages} {061} (\bibinfo {year} {2026})},\ \Eprint
  {http://arxiv.org/abs/2512.19542} {arXiv:2512.19542 [gr-qc]} \BibitemShut
  {NoStop}%
\bibitem [{\citenamefont {Penrose}(1968)}]{Penrose:1968}%
  \BibitemOpen
  \bibfield  {author} {\bibinfo {author} {\bibfnamefont {R.}~\bibnamefont
  {Penrose}},\ }in\ \href@noop {} {\emph {\bibinfo {booktitle} {Battelle
  Rencontres: 1967 Lectures in Mathematics and Physics}}}\ (\bibinfo
  {publisher} {W. A. Benjamin, Inc.},\ \bibinfo {address} {New York},\ \bibinfo
  {year} {1968})\ pp.\ \bibinfo {pages} {121--235}\BibitemShut {NoStop}%
\bibitem [{\citenamefont {Poisson}\ and\ \citenamefont
  {Israel}(1989)}]{Poisson:1989zz}%
  \BibitemOpen
  \bibfield  {author} {\bibinfo {author} {\bibfnamefont {E.}~\bibnamefont
  {Poisson}}\ and\ \bibinfo {author} {\bibfnamefont {W.}~\bibnamefont
  {Israel}},\ }\href {\doibase 10.1103/PhysRevLett.63.1663} {\bibfield
  {journal} {\bibinfo  {journal} {Phys. Rev. Lett.}\ }\textbf {\bibinfo
  {volume} {63}},\ \bibinfo {pages} {1663} (\bibinfo {year}
  {1989})}\BibitemShut {NoStop}%
\bibitem [{\citenamefont {Ori}(1991)}]{Ori:1991zz}%
  \BibitemOpen
  \bibfield  {author} {\bibinfo {author} {\bibfnamefont {A.}~\bibnamefont
  {Ori}},\ }\href {\doibase 10.1103/PhysRevLett.67.789} {\bibfield  {journal}
  {\bibinfo  {journal} {Phys. Rev. Lett.}\ }\textbf {\bibinfo {volume} {67}},\
  \bibinfo {pages} {789} (\bibinfo {year} {1991})}\BibitemShut {NoStop}%
\bibitem [{\citenamefont {Raychaudhuri}(1955)}]{Raychaudhuri:1953yv}%
  \BibitemOpen
  \bibfield  {author} {\bibinfo {author} {\bibfnamefont {A.}~\bibnamefont
  {Raychaudhuri}},\ }\href {\doibase 10.1103/PhysRev.98.1123} {\bibfield
  {journal} {\bibinfo  {journal} {Phys. Rev.}\ }\textbf {\bibinfo {volume}
  {98}},\ \bibinfo {pages} {1123} (\bibinfo {year} {1955})}\BibitemShut
  {NoStop}%
\bibitem [{\citenamefont {Perlick}(2004)}]{Perlick:2004tq}%
  \BibitemOpen
  \bibfield  {author} {\bibinfo {author} {\bibfnamefont {V.}~\bibnamefont
  {Perlick}},\ }\href {\doibase 10.12942/lrr-2004-9} {\bibfield  {journal}
  {\bibinfo  {journal} {Living Rev. Rel.}\ }\textbf {\bibinfo {volume} {7}},\
  \bibinfo {pages} {9} (\bibinfo {year} {2004})}\BibitemShut {NoStop}%
\end{thebibliography}%
\bibliographystyle{apsrev4-1.bst}
%
%
\end{document}